%% file: arXiv_Main.tex
\documentclass[floatfix,aps,pra,superscriptaddress,longbibliography,twocolumn]{revtex4-2}
\usepackage{color}
\usepackage{xcolor}
\usepackage{amsmath}
\usepackage{amsthm}
\usepackage{amssymb}
\usepackage{graphicx}
\usepackage{makecell}
\usepackage{enumitem}
\usepackage{booktabs}
\usepackage{url}
\usepackage{hyperref}

\begin{document}

\title{Expanding functional protein sequence space using high entropy generative models}

\author{Roberto Netti}
\thanks{R.N., E.H. and F.C. contributed equally to this work.}
\affiliation{Sorbonne Universit\'e, CNRS, Department of Computational, Quantitative and Synthetic Biology---CQSB, 75005 Paris, France}

\author{Emily Hinds}
\thanks{R.N., E.H. and F.C. contributed equally to this work.}
\affiliation{Center for Physics of Evolving Systems, University of Chicago, Chicago, IL, USA}
\affiliation{Pritzker School of Molecular Engineering, University of Chicago, Chicago, IL, USA}

\author{Francesco Calvanese}
\thanks{R.N., E.H. and F.C. contributed equally to this work.}
\affiliation{Institut de Physique Th\'eorique, Universit\'e Paris-Saclay, CEA, Gif-sur-Yvette, France}

\author{Rama~Ranganathan}
\email{Corresponding authors. E-mail: ranganathanr@uchicago.edu, martin.weigt@upmc.fr, francesco.zamponi@uniroma1.it}
\affiliation{Center for Physics of Evolving Systems, University of Chicago, Chicago, IL, USA}
\affiliation{Department of Biochemistry and Molecular Biology, University of Chicago, Chicago, IL, USA}

\author{Martin Weigt}
\email{Corresponding authors. E-mail: ranganathanr@uchicago.edu, martin.weigt@upmc.fr, francesco.zamponi@uniroma1.it}
\affiliation{Sorbonne Universit\'e, CNRS, Department of Computational, Quantitative and Synthetic Biology---CQSB, 75005 Paris, France}
\affiliation{Institut Universitaire de France (IUF)}

\author{Francesco Zamponi}
\email{Corresponding authors. E-mail: ranganathanr@uchicago.edu, martin.weigt@upmc.fr, francesco.zamponi@uniroma1.it}
\affiliation{Dipartimento di Fisica, Sapienza Universit\`a di Roma, Piazzale Aldo Moro 5, 00185 Rome, Italy}

\begin{abstract}
Boltzmann Machines trained on evolutionary sequence data have emerged as a powerful paradigm for the data-driven design of artificial proteins. However, the relationship between model architecture, specifically parameter density, and experimental performance remains poorly understood. Here, we investigate this relationship using the Chorismate Mutase enzyme family as a model system. We compare standard fully connected Boltzmann Machines for Direct Coupling Analysis (bmDCA) with sparse models generated via progressive edge activation (eaDCA) and edge decimation (edDCA). We identify a maximum-entropy model (meDCA) along the decimation trajectory that represents an optimal balance between constraint satisfaction and the flexibility of the probability distribution. We synthesized and tested artificial sequences from all models using an \textit{in vivo} complementation assay, finding that all architectures, regardless of sparsity, generate functional enzymes with high success rates, even at significant divergence from natural sequences. Despite this functional equivalence, we demonstrate that the meDCA model samples a viable sequence space that is more than fifteen orders of magnitude larger than its low-entropy counterparts. Furthermore, comparative analyses reveal that high-entropy models systematically minimize overfitting and better capture the local neutral spaces surrounding natural proteins. These findings suggest that while various models satisfying coevolutionary statistics can generate functional sequences, high-entropy Boltzmann Machines provide a superior representation of the underlying evolutionary fitness landscape.
\end{abstract}

\maketitle

\section{Introduction}

\begin{figure*}[t!]
    \centering
    \includegraphics[width=1\linewidth]{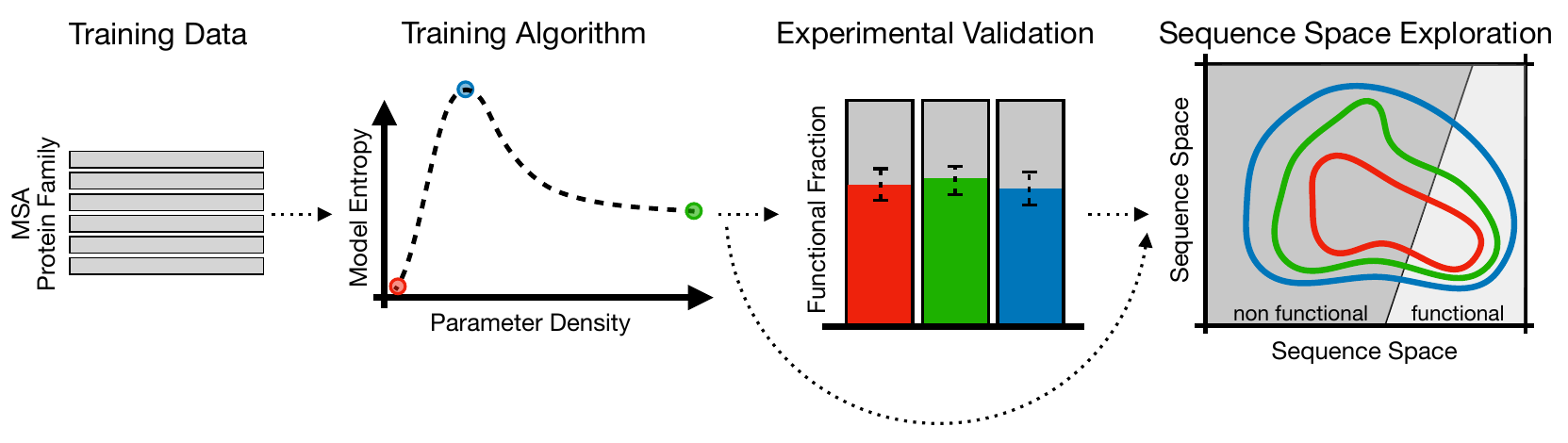}
    \caption{{\bf Pipeline for sparse generative modeling of Chorismate Mutase proteins}. The workflow begins with a multiple sequence alignment (MSA) of natural homologous proteins, which is then used to train a family of generative models. Starting from a fully connected parameterization in which all possible two-site frequencies are used in the inference (green dot), we iteratively remove parameters to induce sparsity. By monitoring the Shannon entropy of the model throughout this process, we observe that entropy initially increases, indicating improved generalization, before decreasing as the model becomes overfitted. We select the maximum-entropy model (blue dot) and a strongly overfitted model (red dot) for comparative analysis. Artificial sequences generated from all three models were tested \textit{in vivo} using a previously validated \textit{E. coli} platform, yielding high and comparable success rates. The maximum-entropy model successfully samples a broader sequence space while maintaining the functional fitness of natural proteins.}
    \label{fig:1}
\end{figure*}

The engineering of novel proteins with specific catalytic activities or structural properties is a central challenge in biotechnology and synthetic biology. While rational design based on physical principles \cite{kuhlman2003design} has made significant progress, data-driven approaches that infer design rules directly from evolutionary history have emerged as a powerful alternative \cite{russ2020evolution, hawkinshooker2021functional, repecka2021proteingan, ferruz2022protgpt2, madani2023large}. By analyzing the statistical patterns embedded in Multiple Sequence Alignments (MSAs) of homologous protein families, generative probabilistic models can learn the constraints specifying biological function and generate synthetic sequences that are statistically indistinguishable from natural ones \cite{figliuzzi2018pairwise, riesselman2018deep, tubiana2019learning, russ2020evolution, trinquier2021efficient}.

Among these approaches, Direct Coupling Analysis (DCA) has proven particularly effective~\cite{russ2020evolution, morcos2011direct, cocco2018inverse}. By constructing a global probability distribution over sequence space that reproduces observed one- and two-site amino acid frequencies, DCA models capture both site-specific conservation and epistatic coevolutionary interactions~\cite{starr2016epistasis, hopf2017mutation, flynn2017inference, chen2024epistatic}. These models define a ``probability landscape'', allowing not only for the sampling of novel variants but also for a deeper understanding of the complex sequence-function relationship by exploring evolutionary constraints and the local geometry of the sequence space surrounding natural and artificial proteins \cite{morcos2014coevolutionary, levy2017potts, figliuzzi2016coevolutionary, poelwijk2019learning, hopf2017mutation, haldane2018coevolutionary}.

However, a fundamental challenge in training these models is the limited size of available sequence data relative to the vast dimensionality of the sequence space. Standard DCA approaches typically infer fully connected interaction graphs (bmDCA), resulting in dense models where the number of parameters often exceeds the information content of the training set. This over-parameterization raises critical questions regarding overfitting and the true topology of the inferred landscape.

Does the inclusion of all possible couplings capture essential biological constraints, or does it introduce noise that artificially constrains the model? Recent computational strategies have proposed sparsification, systematically removing parameters, to filter noise and improve interpretability \cite{barton2016ace, barrat2021sparse, calvanese2024parsimonious}. Yet, the experimental consequences of such sparsity on generative capacity and on the quality of the inferred fitness landscape remain largely unexplored.

In this work, we systematically investigate the relationship between model architecture, Shannon entropy, and generative capability. We focus on the AroQ family of Chorismate Mutase (CM) enzymes \cite{kast1996exploring, szurmant2004divergent}, a system previously established as a benchmark for evolution-based design \cite{russ2020evolution}. We compare the standard fully connected bmDCA with two alternative training procedures, one based on iterative parameter activation (eaDCA) and the other on pruning (edDCA), that result in highly sparse models \cite{barrat2021sparse, calvanese2024parsimonious, rosset2024adabmdca}. By monitoring the evolution of Shannon entropy, we identify a specific regime, termed meDCA, that maximizes the entropy of the model along the decimation trajectory. We validate these models experimentally using an \textit{in vivo} complementation assay, testing their ability to generate functional enzymes across varying degrees of sequence divergence. Our results reveal that while sparse and dense models achieve comparable experimental success rates, maximizing entropy enables the model to access a significantly larger (by more than fifteen orders of magnitude) and more diverse region of the functional sequence space (as schematically shown in Fig.~\ref{fig:1}). Furthermore, we demonstrate that high-entropy models offer a more accurate representation of the neutral space in the vicinity of natural sequences and are systematically more robust to overfitting, significantly assigning higher probability scores to unseen natural sequences.

\section{Results}

\subsection{Why Entropy is Important}

For a family of homologous biological sequences (RNAs or proteins), a generative model aims to accurately infer the ``fitness landscape'' by learning the family's statistical properties and defining a global probability distribution $P(\mathbf{A})$ over the entire space of possible sequences $\mathbf{A} = (a_1,\cdots,a_L)$ of aligned length~$L$~\cite{morcos2011direct, tubiana2019learning, riesselman2018deep, trinquier2021efficient}. A successful model assigns high probability to functional sequences, the ``peaks'' of the landscape, and low probability to non-functional ones, the ``valleys'' of the landscape \cite{weinreich2006darwinian, romero2009exploring}. The quality of a model is determined not only by how well its probability distribution overlaps with the \textit{true} fitness landscape in the vicinity of the training data, but also by how well it generalizes to predict the fitness of unknown or newly generated distant sequences.

We propose that a central concept for evaluating the quality of a generative model is Shannon entropy ${S = -\sum_{\mathbf{A}} P(\mathbf{A}) \log P(\mathbf{A})}$, which has two key interpretations in this context. First, as a \textit{measure of variability}, entropy quantifies the diversity of sequences the model considers viable. High entropy indicates a large variability in the generated samples, meaning the model accesses a larger region of the sequence space. Conversely, low entropy implies that generated sequences will be more similar to one another. This diversity can be quantified by estimating the effective size of the viable sequence space as $\Omega = e^{S}$ (effective number of sequences) \cite{cover1999elements, lambert2024expanding}.

Second, entropy is fundamental to the model construction via the Maximum Entropy Principle (MEP) \cite{cocco2018inverse, morcos2011direct}. This principle, which is the foundation for models like bmDCA, states that the best model is the one that remains maximally ``agnostic'' (makes the fewest assumptions) while reproducing a chosen set of observed evolutionary constraints (e.g. one- and two-site frequencies) from the training data. This approach prevents overfitting the statistical noise inherent in limited training data.

\subsection{Impact of Training Algorithms and Sparsity on Model Entropy}

While the MEP provides the theoretical functional form for the least-constrained model, implementing the training in practice introduces significant complexity. Different training procedures and hyperparameters can produce models that match the target constraints to high accuracy, yet yielding vastly different final entropy values. This work explores how model entropy relates to its generative quality. We hypothesize that, within the DCA framework, higher-entropy models do not merely generate more sequence variability; they generate diverse sequences that more accurately overlap with the true experimental fitness landscape, provided the model is properly trained to achieve high correlation with observed statistics, as we detail below.

To test this idea, we compare models generated by three distinct training algorithms. All models explicitly fit the one-site frequency (conservation) profiles, while they crucially differ in the way they account for two-site (coevolution) profiles, cf.~\textit{Materials and Methods} for technical details. The first model is the standard \textit{fully connected} bmDCA \cite{morcos2011direct, muntoni2021adabmdca}, which retains 100\% of its parameters to fit explicitly all possible pairs of two-site frequencies. The second is a ``top-down'' \textit{edge decimation} approach, called edDCA \cite{barrat2021sparse}. This method starts from a pre-trained, fully connected bmDCA model and iteratively removes the least statistically significant parameters (or ``edges'' in a graphical representation of the model). This choice corresponds to removing a certain subset of the two-site frequencies from those included in the MEP. This procedure reveals a specific entropy evolution upon changing parameter density (sketched in Fig.~\ref{fig:1} and further detailed in Fig.~\ref{fig:entropy_density} below): as parameters are removed, model entropy initially increases, until reaching a peak, beyond which further reduction in parameter density causes an abrupt drop in entropy to values far below that of the initial dense model. The third and last procedure is a ``bottom-up'' \textit{edge activation} (eaDCA) approach starting from an empty graph (profile model), which only fits the one-frequency conservation profile, and then iteratively adding the most significant parameters only, i.e. those offering the maximum likelihood gain \cite{calvanese2024parsimonious}. This method produces highly sparse models characterized by very low entropy, comparable to decimated models at similar low parameter densities (\textit{SI Appendix S1}).

To assess generative quality, we selected four models from the training procedures discussed above, all of comparably high final fitting quality (measured a posteriori by a Pearson correlation of approximately $0.94$ between the full sets of two-site connected correlations, estimated from the original MSA and a large independent sample of artificial sequences generated by the final model):
\begin{itemize}
    \item \textbf{bmDCA:} A fully connected model.
    \item \textbf{eaDCA:} An edge-activated model with only $\sim 3$\% of the parameters of bmDCA.
    \item \textbf{edDCA:} A model obtained by decimation from the bmDCA one, with similar parameter density ($\sim 3$\%).
    \item \textbf{meDCA:} A maximum entropy model identified at the entropy peak along the decimation path of edDCA, with $\sim 12$\% parameter density.
\end{itemize}
We stress that the ``maximum entropy'' designation refers specifically to the peak achieved along our decimation trajectory, rather than the global maximum across all possible DCA models of similar fitting quality, which we cannot explicitly construct. We also stress that the Pearson correlation is calculated for the full matrix of connected two-site correlations, not only on the sparsified graph of non-zero parameters. Despite achieving comparable fitting quality, these models exhibit vastly different entropies, cf.~Table~\ref{tab:dca_comparison}.

It is important to note that the fully connected bmDCA model evaluated here was trained extensively without an early stopping criterion (see \textit{SI Appendix S1}). Because prolonged training continuously decreases model entropy even after the target Pearson score is reached, this tested bmDCA exhibits a low entropy, specifically, lower than the initial dense model used as the starting point for the edDCA process. The entropy gap of $\Delta S \approx 37$ between the most and least entropic models is particularly striking; it implies that the meDCA model samples a sequence space $e^{\Delta S} \approx 10^{16}$ times larger than the sparse eaDCA model (\textit{SI Appendix S2}). This difference underscores how the choice of training algorithm and the specific constraints included in the MEP fundamentally dictate the breadth of the generated landscape.

\begin{table}[t]
    \centering
    \setlength{\tabcolsep}{5pt}
    \renewcommand{\arraystretch}{1.3}
    \begin{tabular}{@{}lcccc@{}}
        \toprule
        \textbf{Model} & eaDCA & bmDCA & edDCA & meDCA \\
        \midrule
        $\rho_{C_{ij}}$ & 0.94 & 0.94 & 0.93 & 0.94 \\
        Parameter density (\%) & 2.79 & 100 & 2.94 & 12.5 \\
        Entropy & 113.35 & 121.82 & 136.05 & 150.13 \\
        \bottomrule
    \end{tabular}
    \caption{Comparison of final fitting accuracy, parameter density and entropy across four DCA models trained on the AroQ family of chorismate mutases (CMs). Reported Pearson correlations are computed a posteriori from independent equilibrium samples generated by the final models, thus providing the final assessment of the model fit; they can be slightly lower than the training Pearson correlations used for stopping during training.}
    \label{tab:dca_comparison}
\end{table}

\subsection{Experimental Test of Generative Capacity on the Chorismate Mutase Enzyme}

\begin{table*}[t]
    \centering
    \setlength{\tabcolsep}{6pt}
    \renewcommand{\arraystretch}{1.3}
    \begin{tabular}{@{}lccc@{}}
        \toprule
        \textbf{Seq divergence} & \textbf{20--25\%} & \textbf{40--45\%} & \textbf{60--65\%} \\
        \midrule
        eaDCA & 31.8\% [26.2\%--38.0\%] & 12.6\% [9.0\%--17.4\%] & 2.0\% [0.9\%--4.6\%] \\
        bmDCA & 28.6\% [23.2\%--34.7\%] & 11.9\% [8.4\%--16.6\%] & 1.3\% [0.4\%--3.7\%] \\
        edDCA & 23.9\% [19.0\%--29.6\%] & 5.2\% [2.8\%--9.6\%] & 0.0\% [0.0\%--4.4\%] \\
        meDCA & 30.9\% [25.5\%--36.9\%] & 5.2\% [3.1\%--8.7\%] & 0.4\% [0.1\%--2.2\%] \\
        \bottomrule
    \end{tabular}
    \caption{Rate of functionality (functional hits over total tested) across sequence divergence intervals. Sequence divergence is given by 100\% minus the percentage sequence identity to the closest natural homolog. Values in brackets represent 95\% confidence intervals calculated using the Wilson score method. The edDCA model experienced severe coverage bottlenecks and reduced sequencing yield; out of 250 designed sequences per batch, only 173 and 84 variants were detected and tested in batches 2 and 3, respectively, potentially depressing its observed functional rates.}
    \label{tab:divergence_comparison}
\end{table*}

To provide a direct experimental benchmark for generative capacity, we used the AroQ family of Chorismate Mutase (CM) enzymes. This family serves as an ideal validation platform, as previous work has confirmed that bmDCA models can successfully generate novel, functional CM enzymes \cite{russ2020evolution}. Sparse models with fewer parameters have, however, never been tested experimentally. We employed a high-throughput \textit{in vivo} complementation assay to quantify the functionality of synthetic sequences generated by each model. As a positive control, the functionality of natural sequences from the training MSA was also assessed (\textit{SI Appendix S3}).

More specifically, we sampled artificial sequences from all four models (eaDCA, bmDCA, edDCA, meDCA) by standard Monte Carlo methods \cite{muntoni2021adabmdca, rosset2024adabmdca}. In order to stringently probe the model ability to generalize beyond known data, sequences were selected within three ranges of sequence divergence from the \textit{closest} (in terms of number of mutated sites) natural homologous sequence in the training alignment:
\begin{itemize}
    \item \textbf{Close:} 20--25\% divergence (75--80\% sequence identity) to closest homolog.
    \item \textbf{Moderate:} 40--45\% divergence (55--60\% sequence identity) to closest homolog.
    \item \textbf{Distant:} 60--65\% divergence (35--40\% sequence identity) to closest homolog.
\end{itemize}

The ``distant'' range represents a stringent test of generalization, sampling a region of sequence space far from any known natural sequence, where the expectation of finding functional sequences is lowest. It should be stressed that most generative models, whatever their architecture, fail to generate viable sequences beyond this distance.

Functional assays were performed on the sequences synthesized from all four models (Table~\ref{tab:divergence_comparison}). As anticipated, every model produced a high proportion of functional sequences within the near-natural range, consistent with established benchmarks for natural homologs. While functionality declined as divergence from the closest natural sequence increased, a measurable fraction of sequences in the moderate and distant ranges still retained comparable activity to the natural ones. Overall, all models exhibited substantial generative capability across the tested divergence ranges, confirming that both sparse and dense models can navigate the sequence landscape and generate novel, functional enzymes. At the same time, the data suggest a modest trend in the more distant regimes: lower-entropy models may achieve slightly higher functional fractions at intermediate and large divergence from the training set. This observation points to a possible trade-off between the functional fraction of sampled sequences far from naturals and the overall entropy, and thus diversity, of the generative model.

\subsection{High-Entropy Models Access Broader Regions of the Functional Sequence Space}

\begin{figure*}[t!]
    \centering
    \includegraphics[width=1\linewidth]{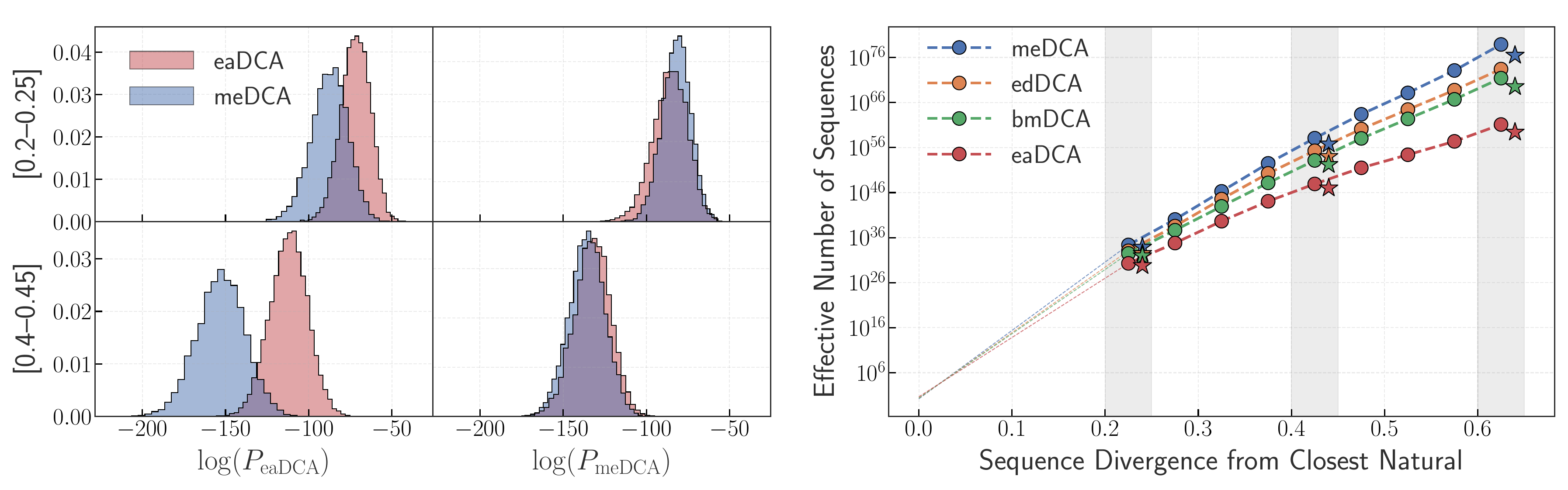}
    \caption{{\bf Assessment of generative capability of DCA models.} \underline{Left panel:} Histograms of statistical energy (log-probability) distributions assigned by a ``scoring'' model (columns) to sequences generated by a ``generating'' model (legend). Sequences are taken from the \textit{close} (0.2--0.25) and \textit{moderate} (0.4--0.45) divergence bins (rows). (Left column) The low-entropy eaDCA model assigns low probability to sequences from the meDCA model. (Right column) The high-entropy meDCA model assigns comparable probability to sequences from the eaDCA model, demonstrating its broader coverage. \underline{Right panel:} The plot shows the entropy (represented as the effective number of sequences $\Omega =\exp S$) of each model, binned by divergence from the closest natural homolog. At all divergence levels, the meDCA model (blue) generates a substantially more diverse set of sequences than the other models. Stars identify the estimated number of functional sequences accessible by each model (i.e., the experimental functional rate multiplied by $\Omega$) in each bin of divergence from the naturals.}
    \label{fig:cross_model_scoring}
\end{figure*}

The experimental results in Table~\ref{tab:divergence_comparison} confirm that all four models, regardless of entropy or parameter density, possess comparable generative capacity. They successfully generalize to discover novel, functional CMs even at significant sequence divergence. This raises a critical question: if all models are generatively comparable, what is the practical implication of the large differences in model entropy?

We hypothesized that while all models can find functional sequences, high-entropy models identify a broader, more diverse set of them. To test this, we performed a cross-model scoring analysis, computing the log-probability (statistical energy) of sequences generated by one model using the score of another. This analysis reveals a distinct asymmetry, best illustrated by comparing the lowest-entropy model (eaDCA) and the highest-entropy model (meDCA). As shown in Fig.~\ref{fig:cross_model_scoring}, the low-entropy eaDCA model assigns a poor probability (compared to its own sample) to a large fraction of the sequences generated by meDCA. Conversely, the meDCA model assigns high probability (same as its own sequences) to those generated by eaDCA. This indicates that the eaDCA model defines a narrow region of the sequence landscape and misses subregions identified by the meDCA model. On the contrary, the meDCA model generates a broader landscape that \textit{encompasses} the space defined by the lower-entropy model. This finding is general across all model pairings and becomes more pronounced as the entropy difference increases (\textit{SI Appendix Fig. S2}).

This broader coverage translates directly into a more diverse output of generated sequences. We quantified this by calculating the entropy of each model within each divergence bin, expressed as the effective number of sequences $\Omega = e^{S}$ \cite{lambert2024expanding}. Fig.~\ref{fig:cross_model_scoring} shows that at every divergence level, the meDCA model consistently explores a larger effective number of sequences than the other models. Crucially, this significantly greater diversity is achieved despite a modest reduction in functional fraction in the more distant sequence regimes. Because the sampling strategy was uniform, the experimentally observed functional rate acts as an estimator of the functional density within the total generable space of each model. The data reveal a critical difference: while the functional \textit{rates} are comparable, the total effective number of generable sequences ($\Omega$) differs by orders of magnitude. Consequently, the \textit{estimated number} of functional sequences that the meDCA model identifies (functional rate $\times$ $\Omega$) remains orders of magnitude larger. Taken together, these results demonstrate that while there may be a mild trade-off between functional fraction far from the training set and overall entropy, this effect is overwhelmingly dominated by the much broader region of sequence space accessed by the meDCA model. As a result, the meDCA model (selected at the entropy peak of the decimation procedure) provides the most comprehensive exploration of the functional landscape, accessing a wider and more diverse region of the viable sequence space.

\subsection{High-Entropy Models Minimize Overfitting to the Training Data}

\begin{figure*}[t!]
    \centering
    \includegraphics[width=1\linewidth]{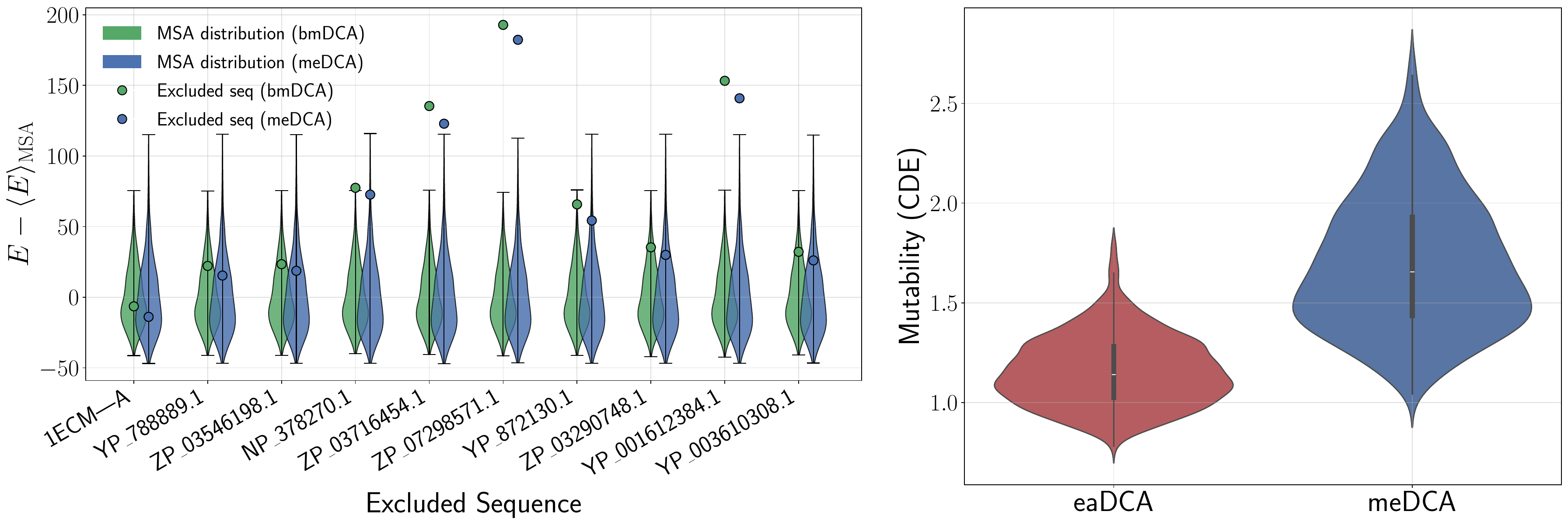}
    \caption{\underline{Left panel}: \textbf{Assessment of generalization capability on unobserved natural sequences.} To evaluate the ability of the models to generalize to regions of sequence space not covered by the training data, we perform a leave-one-group-out validation test. We selected 10 mutually distant natural sequences ($>80\%$ divergence). For each target, we constructed a specific training set by excluding the target sequence and all its close homologs ($<20\%$ divergence), and we used it to train the bmDCA and meDCA models. The horizontal axis in the plot displays the label of each of the 10 excluded sequences. The vertical axis displays the statistical energy assigned by the two models, centered by subtracting the mean energy of the natural sequences. The meDCA model systematically assigns a lower relative energy (and thus a higher probability) to the unobserved sequence compared to the fully-connected bmDCA model. This result indicates that the meDCA model is less prone to overfitting and has better capacity to recognize functional sequences distant from the training data. \underline{Right panel}: \textbf{Mutability distribution of natural Chorismate Mutase sequences.} Mutability is quantified as the site-specific context-dependent entropy averaged over the $L$ sites of each sequence in the training MSA. While the low-entropy eaDCA model assigns low mutability scores to natural sequences, the high-entropy meDCA model assigns significantly higher values. This broader mutability profile is consistent with experimental evidence of a large neutral mutational space surrounding wild-type sequences, indicating that the high-entropy model more accurately captures the local permissiveness of the fitness landscape (\textit{SI Appendix Fig. S6}).}
    \label{fig:ene}
\end{figure*}

Standard fully connected DCA models often struggle with high dimensionality, because their parameter space is large relative to the limited size of available training sequences~\cite{barrat2021sparse, muntoni2021adabmdca, rosset2024adabmdca, barton2016ace}. While the MEP is designed to construct the least biased model consistent with observed statistics, the inclusion of noise-induced two-site frequencies, lacking statistical significance, can introduce spurious couplings and lead to overfitting. Sparsification, or the systematic removal of these couplings, offers a pathway to mitigate this effect. However, as demonstrated in the previous sections, excessive parameter removal leads to a sharp reduction in model entropy. The few remaining couplings must be extremely overfitted to large values to reach the overall fitting criterion, which concerns all pairwise covariances, cf. above. We hypothesize that the meDCA model, which maximizes entropy along the decimation trajectory, represents an optimal balance: here, parameter reduction effectively filters out noise without compromising global constraints, thereby maximizing generalization capability.

Quantifying overfitting in generative models is non-trivial, but generally overfitted models provide far higher losses (or here statistical energies) to validation data outside the training set than to the training data. To rigorously assess this, we performed a \textit{leave-one-group-out} validation strategy. We selected 10 mutually distant natural sequences ($>80\%$ divergence) and constructed 10 corresponding training sets, each formed by excluding one target sequence along with all its close neighbors ($<20\%$ divergence) (\textit{SI Appendix Fig.~S3, S4}). This procedure effectively creates a distinct \textit{hole} in the training data around the target. We subsequently trained both bmDCA and meDCA on these reduced datasets and computed the statistical energy of the unseen target sequences. The comparison reveals a systematic trend: the meDCA model assigns energies to the withheld sequences that are more comparable to the distribution of the training data than those assigned by the dense bmDCA model. The dense model is effectively more \textit{surprised} by unseen valid sequences (Fig.~\ref{fig:ene}, \textit{SI Appendix Fig.~S5}). This indicates that the high-entropy model maintains a more comprehensive representation of the fitness landscape, yielding a small but systematic gain in generalization power for sequences far removed from the training set.

\subsection{Representation of Local Neutral Spaces}

A direct consequence of overfitting is the construction of a rugged landscape where probability peaks are sharp and isolated around the training sequences. In an overfitted model, the probability landscape does not interpolate well between data points; even sequences that are one or two mutations away from a training sequence are often assigned drastically lower probabilities, implying a rigid intolerance to mutation. However, experimental evolution and deep mutational scanning studies \cite{fowler2014deep} have repeatedly demonstrated that natural proteins reside within extensive \textit{neutral spaces}, regions of the sequence landscape where mutational variation preserves function, cf.~e.g.~\cite{jacquier2013tem1, stiffler2015evolvability, rockah2015mapping, tokuriki2009stability}. Natural sequences must then be highly mutable, in order to permit the evolutionary navigation of this neutral space. To quantify whether our models capture this biological reality, we analyzed the average Context-Dependent Entropy (CDE)~\cite{vigue2022deciphering, rodriguez2022epistatic}. This metric is defined as the entropy of the conditional probability distribution of a single residue given the context of all other residues in the sequence, averaged over all sites, cf.~\textit{Materials and Methods} for a proper definition. A high average CDE indicates that many residues are free to mutate without disrupting the global statistical compatibility of the sequence, while a low CDE signals a small number of allowable mutations. The average CDE effectively serves as a proxy for the mutability of the protein.

When computing the average CDE for natural sequences within the training MSA, a striking difference emerges: the meDCA model assigns significantly higher mutability scores compared to its lowest-entropy counterpart eaDCA (Fig.~\ref{fig:ene}). Given that our experimental validation confirmed the high functional yield of meDCA in the vicinity of natural sequences, this increased local entropy cannot be dismissed as noise or an artifact of loose constraints. Rather, it could indicate that the meDCA model does not merely memorize the training sequences but interpolates the neutral space between them. By smoothing the landscape around the probability peaks related to sequences from the training set, the high-entropy model provides a more realistic representation of the local permissiveness of the fitness landscape, thereby facilitating a more effective navigation and exploration of the sequence space \cite{mauri2023transition} (\textit{SI Appendix Fig.~S6, S7}). This enhanced explorability is directly reflected in the sampling dynamics: the meDCA model exhibits a dramatically faster mixing time compared to lower-entropy variants, mixing approximately 2, 3, and 40 times faster than edDCA, eaDCA, and the fully connected bmDCA, respectively (\textit{SI Appendix Fig.~S7}). Such a significant acceleration indicates a vastly superior navigability of the sequence landscape, where energetic barriers separating distinct high-probability clusters are smoother and easier to traverse.

\section{Discussion}

\subsection{Generative Equivalence versus Landscape Breadth}

We initiated this study to determine if different training algorithms, yielding generative models in the DCA class but with vastly different parameter densities and entropies, provide distinct insights into the fitness landscape of a protein family. By systematically comparing four Boltzmann Machine variants, ranging from fully connected to highly sparse, we arrived at a counter-intuitive result: all models proved to be effective generators. Consistent with previous benchmarks, every model successfully generated functional Chorismate Mutase enzymes, even at significant sequence divergence from natural homologs. This confirms that the fundamental evolutionary constraints captured by DCA are sufficient for generative design, regardless of the specific training strategy employed \cite{russ2020evolution}.

However, this comparable generative accuracy masks a critical disparity in the \textit{quality} and \textit{coverage} of the modeled landscapes. Our cross-model scoring analysis revealed a hierarchical relationship between the models: high-entropy models generate a broad landscape that statistically encompasses the narrower regions captured by lower-entropy models. Specifically, the very sparse eaDCA model assigns low probabilities to valid sequences generated by the meDCA model, indicating that it does not recognize these sequences as part of its own fitness landscape. The reverse is not true: the meDCA model assigns similar probability to its own sequences and to those generated by the eaDCA model, indicating that it considers the landscape of the eaDCA model as a subset of its own. This asymmetry demonstrates that lower-entropy models tend to overfit specific parts of the landscape, effectively ``missing'' functional subregions. Note that this is different from mode collapse observed in some generative model architectures, since it is rather related to landscape roughness rather than to the lack of entire subfamilies. By selecting the model at the entropy peak of the decimation process, we avoid this overfitting and access an effective \textit{functional} sequence space $\Omega$ that is orders of magnitude larger than that of the other models. Although our experimental data suggest that lower-entropy models may retain a slightly higher functional fraction in the far-from-natural regime, this effect remains minor relative to the enormous gain in accessible diversity.

\subsection{Implications for Model Selection}

These findings establish a clear criterion for model selection based on specific scientific or engineering objectives. If you just need a functional sequence, any well-trained DCA model can work. But for advanced applications like de novo design, high-entropy models offer a major advantage: they greatly expand the range of accessible sequences and can reach functional regions far from any known natural homolog. In addition, for scientific inquiries aiming to use the model as a proxy to study the evolutionary landscape itself, model choice is paramount. A low-entropy model provides a statistically narrow and incomplete map by overfitting to specific probability peaks. To capture the full diversity of sequence-function relationships and the subtle constraints shaping evolution, the high-entropy model, tuned via our parameter-density modulation, is essential. Consequently, the decimation procedure presented here should be viewed not merely as a compression technique, but as a tool to achieve a comprehensive representation of the functional landscape. Furthermore, the smoother barriers characteristic of the meDCA landscape drastically reduce the mixing time facilitating a much faster and smoother navigation of a highly connected functional sequence space \cite{mauri2023transition, greenbury2022navigable, wu2016adaptation, poelwijk2007accessible}, with the consequent practical advantage of generating independent equilibrium sequences at a proportionally faster rate.

\subsection{Future Perspectives}

A critical next step is to extend this evaluation to alternative model architectures, including deep generative and protein language models. As an initial step in this direction, we trained a small set of autoregressive decoder-only Transformer models on the same Chorismate Mutase family. The resulting comparison indicates that, in this limited-data regime, Transformer entropy depends strongly on the stopping criterion. When the stopping epoch is selected by validation loss, the models preserve entropies comparable to those of DCA models, but their agreement with the natural pairwise correlations $C_{ij}$ and the general fitting quality remain substantially lower. In contrast, when training is continued until the agreement with $C_{ij}$ becomes comparable to that of the DCA models, the entropy of the Transformer models drops sharply (\textit{SI Appendix Table S1}). Although these Transformer-generated sequences were not experimentally tested in the present work, this preliminary analysis suggests that, in single-family settings, especially when training data are limited, DCA models retain a favorable balance between fitting coevolutionary statistics and preserving landscape breadth, and may therefore offer a practical advantage when the goal is to maximize the accessible sequence space. This observation further motivates a broader comparison with modern generative frameworks. These architectures, often trained on massive, unaligned, and multi-family datasets, represent a paradigm shift in generative modeling \cite{rives2021biological, ferruz2022protgpt2, madani2023large, repecka2021proteingan}. While the specific analytical framework of this study may require adaptation, the core philosophy, testing generalization far from training data and quantifying the breadth of the mapped landscape, remains essential to validate their utility in biological design.

\section{Materials and Methods}

\subsection{Training algorithms for sparse Boltzmann Machines}

\subsubsection*{Short Introduction to Direct Coupling Analysis}

The core methodological framework employed in this study is Direct Coupling Analysis (DCA). DCA is a statistical modeling approach used to infer evolutionary constraints from Multiple Sequence Alignments (MSAs) of homologous protein families. Conceptually, the method constructs a global probability distribution over the sequence space. This distribution is constrained to match the empirical one-site and two-site amino acid frequencies observed in the MSA, while otherwise remaining as unconstrained as possible, according to the Maximum Entropy Principle \cite{cocco2018inverse, morcos2011direct}. The resulting model takes the form of a Potts model in statistical physics language (or Boltzmann Machine in machine learning language):
\begin{equation}
P(\mathbf{A}) = \frac{1}{Z} \exp \left( \sum_{i} h_i(a_i) + \sum_{i<j} J_{ij}(a_i, a_j)\right) .
\label{eqn:potts_proba}
\end{equation}
where $\mathbf{A} = (a_1, \dots, a_L)$ represents an amino acid sequence of length $L$, $Z$ is the partition function, $h_i(a_i)$ represent local fields describing site-specific conservation, and $J_{ij}(a_i, a_j)$ are pairwise couplings capturing coevolutionary correlations between sites $i$ and $j$. These parameters are typically inferred via gradient ascent on the log-likelihood, ensuring that the model marginals reproduce the empirical statistics. Standard DCA infers a fully connected interaction graph, resulting in a dense model containing a vast number of $q^2L(L-1)/2$ parameters (\textit{SI Appendix S1A}). Many of these couplings are statistically insignificant or arise from noise, potentially leading to overfitting. To address this, sparsification strategies have been previously developed to filter out noise and identify essential constraints. In this work, alongside the standard fully connected model, we employed specific adaptations of these strategies to train three alternative models. This comparative approach allowed us to systematically investigate the impact of training algorithms and model entropy on generative performance.

\subsubsection*{Pruning Algorithm (edDCA)}

\begin{figure}[t]
    \centering
    \includegraphics[width=1\linewidth]{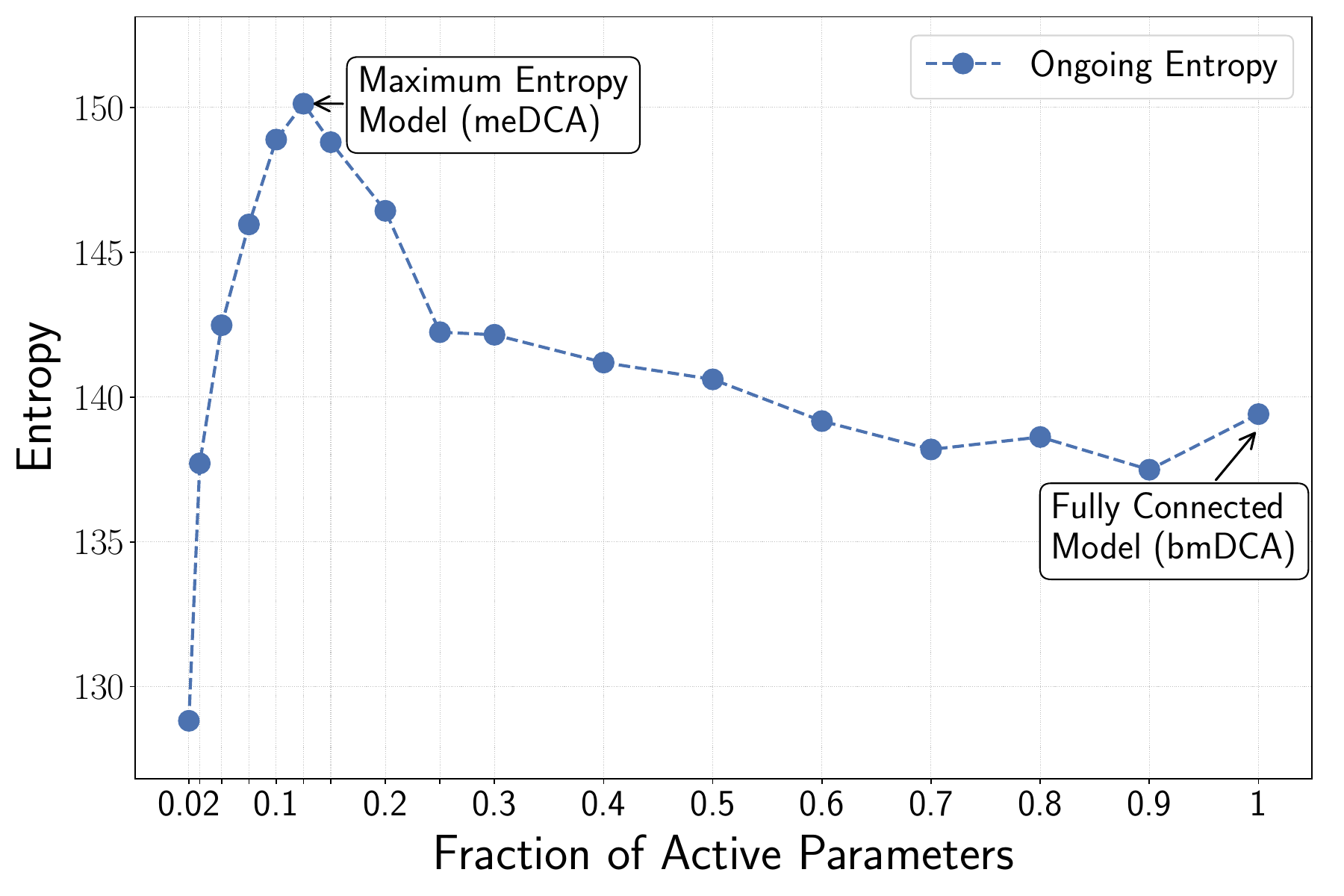}
    \caption{\textbf{Evolution of model entropy as a function of parameter density during the iterative pruning process.} The plot displays the model entropy (y-axis) against the fraction of active parameters (x-axis). The trajectory moves from right to left, starting from the fully connected bmDCA model (density =1.0). The procedure reveals a non-monotonic behavior: initially, pruning weak or noisy parameters relaxes the distribution, leading to an increase in entropy. This trend continues until a peak is reached at a density of approximately $12.5\%$, representing the maximally agnostic model. Beyond this point, further sparsification forces the remaining parameters to become overly rigid to maintain the global Pearson correlation threshold ($\rho>0.94$), resulting in a sharp collapse of the accessible sequence space (\textit{SI Appendix Fig. S8}).}
    \label{fig:entropy_density}
\end{figure}

To systematically identify the entropy peak, we employed an iterative pruning algorithm adapted from Barrat-Charlaix et al. \cite{barrat2021sparse}, which allows us to monitor the dynamics of model entropy as a function of parameter density. Specifically, we first trained a fully connected Boltzmann machine (bmDCA) until the Pearson correlation between the empirical two-point connected correlations and those estimated from the persistent Monte Carlo chains reached $\rho = 0.95$. We refer to this quantity as the training Pearson correlation, to distinguish it from the final a posteriori estimate reported in Table~\ref{tab:dca_comparison}. Subsequently, the algorithm proceeds through an iterative decimation procedure. At each step, couplings $(i,j,a,b)$ are ranked by their statistical significance, quantified by the symmetric Kullback-Leibler divergence between the model with the coupling active versus inactive:
\begin{equation}
D^\mathrm{prune}_{(i,j,a,b)} =
D_{\text{KL}}\left(
P_{(i,j,a,b)\in\mathcal{E}}
\;\|\;
P_{(i,j,a,b)\notin\mathcal{E}}
\right) ,
\label{eqn:pruning_dkl}
\end{equation}
where $\mathcal{E}$ represents the ensemble of active links. We permanently remove the bottom $0.1\%$ of these couplings by permanently setting $J_{ij}(a,b) = 0$. Following each pruning step, we perform Monte Carlo sampling and gradient updates on the surviving parameters to recover the statistical fit, repeating this process to ensure that the training Pearson correlation does not fall below the minimum threshold of $\rho = 0.94$ (\textit{SI Appendix S1B}). This procedure reveals a non-monotonic behavior in model entropy (Fig.~\ref{fig:entropy_density}). Initially, a reduction in parameter density leads to an increase in entropy. This trend arises because the dense model includes numerous weak or meaningless parameters that excessively constrain the sequence space; their removal effectively relaxes the distribution. Crucially, during this initial phase, the model requires no updates to the remaining parameters to maintain its statistical fit. The entropy subsequently reaches a distinct peak at a parameter density of approximately $12.5\%$, beyond which it drops abruptly. This collapse happens because the model retains too few explicit constraints to naturally maintain this global fitting threshold. Consequently, the remaining parameters are forced to fit their specific targets almost perfectly (with a correlation on fitted constraints of $\approx 0.99$). This extreme rigidity severely restricts the accessible sequence space, thereby reducing the total model entropy (\textit{SI Appendix S1C}).

\subsubsection*{Element Activation Algorithm (eaDCA)}

The second method used to explore the low-density regime is an adaptation of the Edge Activation DCA (eaDCA) method \cite{calvanese2024parsimonious}. While standard eaDCA typically activates full coupling matrices (edges) for position pairs $(i,j)$, we aimed to reduce parameters further by activating individual interaction elements $(i,j,a,b)$, similar to the pruning approach. Starting from a profile model (containing only fields), the algorithm iteratively adds parameters to the model. At each step, we identify the most significant position pair $(i, j)$ that maximizes the Kullback-Leibler divergence between the empirical and model two-site marginals:
\begin{equation}
D^\mathrm{activate}_{(i,j)} = D_{\text{KL}}(p_{ij}\| f_{ij}) .
\label{eqn:activation_dkl}
\end{equation}
Once the pair $(i, j)$ is selected, we rank the potential contribution of all specific amino acid interactions $(a, b)$ within that pair. We then activate only the single interaction element $J_{ij}(a,b)$ that, given the selected $(i,j)$, maximizes $D_{\text{KL}}(p_{ij}(a,b)|| f_{ij}(a,b))$. The value of this new parameter is assigned optimally based on the ratio of empirical to model probabilities:
\begin{equation}
\Delta J_{ij}(a,b) = \log \left( \frac{f_{ij}(a,b)(1-p_{ij}(a,b))}{p_{ij}(a,b)(1-f_{ij}(a,b))}\right),
\label{eqn:activation_formula}
\end{equation}
for specific $(i, j, a, b) \in \mathcal{E}$. To ensure numerical stability and regularization, we applied pseudo-counts ($\lambda = 0.8$) to the frequencies during the update step \cite{morcos2011direct, cocco2018inverse}. Following each addition, the algorithm performs MCMC sampling to estimate the new two-point probabilities and identify the next target coupling. Crucially, the selection process considers both inactive and currently active parameters; if an active parameter is selected, its value is simply adjusted to maximize the likelihood rather than adding a new edge. The procedure terminates once the training fitting accuracy reaches a threshold of $\rho=0.95$, resulting in a highly sparse model composed solely of the most information-rich interaction elements (\textit{SI Appendix S1D}).

\subsection{Data Generation and Experimental Validation}

\subsubsection*{Training data and artificial generation}

For the training of our models, we utilized the Multiple Sequence Alignment (MSA) of the AroQ family of Chorismate Mutase (CM) enzymes described in Russ et al. \cite{russ2020evolution}. This dataset comprises 1,259 natural homologous domains, representing a broad diversity of bacterial, archaeal, fungal, and plant lineages. All four generative models defined in this study, dense bmDCA, eaDCA, edDCA, and meDCA, were trained on this specific alignment using the algorithms detailed in the previous section. Following training, we employed each model to generate a large pool of synthetic sequences via Monte Carlo sampling (\textit{SI Appendix S2}). To rigorously test the generalization capability of the models, these synthetic sequences were stratified into three distinct ranges of sequence divergence relative to the closest natural homolog in the training set: close (20--25\%), moderate (40--45\%), and distant (60--65\%). For every sequence in each divergence bin, we computed the statistical energy using the Hamiltonian of the generating model. To curate the final library for experimental testing, we restricted our selection to the top $10\%$ of sequences with the lowest statistical energy (highest probability) within each bin. From this high-probability pool, we uniformly sampled 250 sequences at random for each model and divergence range.

\subsubsection*{Context-dependent entropy and mutability}

To quantify the local mutational permissiveness of a sequence under a given generative model, we computed the context-dependent entropy (CDE). For a sequence $\mathbf{A}=(a_1,\ldots,a_L)$ and a site $i$, we considered the conditional distribution assigned by the model to all possible amino acids at that position while keeping the remaining sequence context fixed, $\mathbf{A}_{\setminus i}=(a_1,\ldots,a_{i-1},a_{i+1},\ldots,a_L)$. The site-specific CDE is defined as
\begin{equation}
    \mathrm{CDE}_i(\mathbf{A}_{\setminus i})
    =
    - \sum_a P(a_i=a \mid \mathbf{A}_{\setminus i})
    \log P(a_i=a \mid \mathbf{A}_{\setminus i}) .
\end{equation}
Unlike entropy measures based only on site-wise amino-acid frequencies (Context Independent Entropy), the CDE depends explicitly on the full sequence background and therefore measures how many substitutions are locally compatible with the global statistical constraints learned by the model. We then summarized the mutability of an entire sequence by averaging this quantity over all aligned positions,
\begin{equation}
    \overline{\mathrm{CDE}}(\mathbf{A})
    =
    \frac{1}{L}
    \sum_{i=1}^{L}
    \mathrm{CDE}_i(\mathbf{A}_{\setminus i}) .
\end{equation}
High values of $\overline{\mathrm{CDE}}$ indicate that, on average, many sites retain a broad conditional distribution of allowed amino acids in their native sequence context, consistent with a locally permissive neutral space. Low values instead indicate that the model assigns only a small number of acceptable substitutions around the sequence, corresponding to a sharper and more constrained local probability landscape~\cite{vigue2022deciphering, rodriguez2022epistatic} (\textit{SI Appendix S4}).

\subsubsection*{Transformer entropy comparison}
As an additional comparison, we trained a small set of autoregressive decoder-only Transformer models on the same Chorismate Mutase MSA. For each architecture, we considered two stopping criteria. In the first case, a validation split was used to identify the epoch at which the validation loss reached its minimum; the model was then retrained from scratch on the full dataset up to this selected epoch. This validation-based stopping criterion preserved relatively high entropy, but left the model substantially underfitted with respect to the training distribution, as reflected by the low agreement with the natural pairwise correlations $C_{ij}$ and by a training loss of approximately $1.5$ per residue. In the second case, training was monitored by periodically sampling sequences from the model, computing the pairwise correlations $C_{ij}$ on the generated sample, and comparing them with those of the training MSA. For comparison with the final fitting quality of the DCA models reported in Table~\ref{tab:dca_comparison}, training was stopped when the Pearson correlation between natural $C_{ij}$ and those estimated from equilibrium samples generated by the current model exceeded $0.94$. This correlation-based stopping criterion brought the Transformer models to a DCA-like level of pairwise-statistics fitting, but was associated with a sharp entropy collapse, reaching values below $30$, consistent with strong overfitting. We varied the embedding size, number of layers, and number of attention heads, and estimated the entropy of the resulting models, as summarized in \textit{SI Appendix Table S1}.

\subsubsection*{Experimental procedure}

The functional characterization of the designed sequences was performed using a high-throughput \textit{in vivo} complementation assay, adapted from the protocol established by Russ et al. \cite{russ2020evolution}. Designed sequences were synthesized as oligonucleotides and cloned into the pKTCTET-0 expression vector. This library was then transformed into the chorismate mutase-deficient \textit{E. coli} strain KA12/pKIMP-UAUC, which is auxotrophic for tyrosine and phenylalanine. To perform the selection, the transformed population was inoculated into M9c selective minimal media lacking tyrosine and phenylalanine, and supplemented with doxycycline to induce gene expression. Under these restrictive conditions, bacterial growth rates are quantitatively coupled to the catalytic activity of the expressed CM variant. Following a 24-hour growth period at 30$^\circ$C, plasmids were purified to serve as the selected sample. To assess functionality, we performed Illumina MiSeq deep sequencing on the library population both before (input) and after (selected) the growth period. The functional performance of each variant was quantified by its relative enrichment ($r.e.$), calculated as the logarithm of the frequency ratio of the variant in the selected population versus the input population, and normalized against the wild-type \textit{E. coli} CM reference. Finally, variants were classified as functional if their normalized enrichment scores fell more than three standard deviations above the mean of the near-null mode, determined by fitting a two-mode Gaussian mixture model (\textit{SI Appendix S3 and Fig. S1}).

\medskip
\noindent
{\bf Data, Materials, and Software Availability -}
All data, processed experimental results, and custom scripts used for the data analysis are publicly available on GitHub at \url{https://github.com/robertonetti/entropy-sparse-DCA}.
The general software package used to train the bmDCA, eaDCA, and edDCA generative models is open-source and available at \url{https://github.com/spqb/adabmDCA} \cite{rosset2024adabmdca}.
The custom code pipeline specifically developed to train the maximum-entropy model (meDCA) and compute the corresponding thermodynamic integration is available at \url{https://github.com/robertonetti/highentropyDCA}.

\medskip
\noindent
{\bf Acknowledgments -}
We thank P.~Barrat-Charlaix, A.~P.~Muntoni, L.~Rosset and K.~S.~Shimagaki for many useful discussions related to the construction of sparse DCA models. We also acknowledge financial support from the Horizon Europe MSCA Staff Exchange project ``SIMBAD'' (grant agreement no. 101131463).

\medskip
\noindent
{\bf Author Contributions -}
All authors designed the research, analyzed the data and wrote the manuscript. R.N. and F.C. implemented the computational methods and generated the sequences for experimental testing. E.H. performed the experiments.

\bibliography{refs}{}

\input{arXiv_Supplementary_Information.tex}

\end{document}

%% file: arXiv_Supplementary_Information.tex
\clearpage
\onecolumngrid

\renewcommand{\thesection}{S\arabic{section}}
\renewcommand{\thefigure}{S\arabic{figure}}
\renewcommand{\thetable}{S\arabic{table}}
\setcounter{section}{0}
\setcounter{figure}{0}
\setcounter{table}{0}

\section*{Supporting Information}
\section{Sparse generative models of protein sequences}
\label{SI:sparse_models}

We describe here the training algorithms for the fully connected Boltzmann Machine (bmDCA) and for the sparse generative models used in the main text.

\subsection{Direct Coupling Analysis and bmDCA}
\label{SI:bmDCA}
   
The Potts model \cite{cocco2018inverse, morcos2011direct} defines the statistical energy of an amino acid sequence $\mathbf{A} = (a_1, \dots, a_L)$. Each position $a_i$ takes a value from a vocabulary of $q=21$ symbols (the 20 standard amino acids plus a gap symbol). The energy is expressed as follows:
\begin{equation}\label{eq:prob_DCA}
    E(\mathbf{A}) = \sum_{(i,a)\in \mathcal{V}} h_i(a) \delta_{a_i,a} + \sum_{(i,a,j,b)\in \mathcal{E}} J_{ij}(a, b) \delta_{a_i,a} \delta_{a_j,b}
\end{equation}
This formulation can be represented on an interaction graph $\mathcal{G}=(\mathcal{V},\mathcal{E})$ using a one-hot encoding format. The vertices $\mathcal{V}$ represent all possible symbol-site combinations, indexed by $(i,a) \in \{1,\dots,L\}\times \{1,\dots,q\}$. The edges $\mathcal{E}$ connect pairs of vertices $(i,a)$ and $(j,b)$. 
The first term in the energy equation utilizes the local fields $h_i(a)$, which capture site-specific conservation patterns. The second term utilizes the pairwise couplings $J_{ij}(a,b)$, which capture epistatic coevolutionary interactions between amino acids at positions $i$ and $j$ \cite{morcos2011direct,cocco2018inverse}. From this statistical energy, the probability of a sequence is defined by the Boltzmann distribution, $P(\mathbf{A}) = e^{-E(\mathbf{A})}/Z$, where the partition function $Z$ normalizes the distribution across all possible sequences.
The model is trained to maximize the log-likelihood of the sequence dataset. For a fully connected Direct Coupling Analysis model (bmDCA), the interaction graph $\mathcal{G}$ includes all possible coupling terms. We define the model parameter density, $d$, as the fraction of non-zero couplings ($J_{ij}(a,b) \neq 0$) relative to the total number of possible couplings, $q^2L(L-1)/2$. Here, density is calculated element-wise for each specific interaction $(i,j,a,b)$, rather than block-wise for the entire $q \times q$ coupling matrix between sites $i$ and $j$  \cite{calvanese2024parsimonious}. 

\subsubsection*{Training Algorithm} The parameters are updated by gradient ascent to ensure the model marginals reproduce the empirical statistics: 
\begin{equation}\label{eq:params_update}
    h_i(a) \leftarrow h_i(a) + \gamma (f_{i}(a) - p_i(a)) \ , \qquad J_{i j}(a, b) \leftarrow J_{i j}(a, b) + \gamma (f_{ij}(a, b) - p_{ij}(a,b))
\end{equation}
where $\gamma$ is the learning rate, $f_i(a)$ and $f_{ij}(a,b)$ are the empirical frequencies from the alignment, and $p_i(a) = \langle \delta_{a_i,a}\rangle$ and $p_{ij}(a, b)=\langle \delta_{a_i,a}\delta_{a_j,b}\rangle$ are the model's marginal probabilities. Training converges when the model marginals match the empirical data frequencies  \cite{muntoni2021adabmdca, rosset2024adabmdca}.
Because computing the partition function $Z$ exactly is intractable, we cannot analytically determine the model probabilities $p_i(a)$ and $p_{ij}(a,b)$. We therefore use Persistent Contrastive Divergence \cite{tieleman2008training}, splitting each training epoch into two distinct steps: 
\begin{itemize}
    \item \textbf{Monte Carlo Sampling:} We simulate independent sequences from the current probability distribution to estimate the one- and two-site marginals at each epoch.
    \item \textbf{Parameter Update:} We adjust the fields and couplings using the gradient ascent rule described above.
\end{itemize}
To determine when to terminate training, we monitor the two-site connected correlation functions for both the empirical data and the model:
\begin{equation}
    C^{\mathrm{data}}_{ij}(a, b) = f_{ij}(a, b) - f_i(a) f_j(b) \ ,\qquad \qquad C^{\mathrm{model}}_{ij}(a, b) = p_{ij}(a, b) - p_i(a) p_j(b)
\end{equation}
Training stops when the Pearson correlation coefficient between the model and empirical connected correlations reaches a predefined target value, empirically set to $0.95$ based on previous benchmarks \cite{muntoni2021adabmdca, rosset2024adabmdca, calvanese2024parsimonious}.

\subsection{Edge Decimation Algorithm (edDCA)}
\label{SI:edDCA}

Edge Decimation (edDCA) is a top-down sparsification procedure \cite{barrat2021sparse}. This approach starts from a fully connected bmDCA model that has already been pre-trained to the target Pearson correlation threshold. The decimation relies on an information-driven algorithm: at each step, we permanently fix specific couplings to zero ($J_{ij}(a,b) = 0$), effectively removing the corresponding interaction edges from the graph $\mathcal{E}$. Once removed, these parameters are excluded from further updates, and the model is no longer explicitly constrained to fit their corresponding empirical two-site frequencies $f_{ij}(a,b)$.
The effect of removing a coupling $J_{ij}(a,b)$ is quantified by the symmetric Kullback-Leibler (KL) divergence between the Boltzmann distributions with and without that coupling \cite{barrat2021sparse}. Let $E(\mathbf{A})$ be the statistical energy of the full Potts model, and $E'(\mathbf{A})$ be the energy of the model where the coupling has been removed:
\begin{equation}
    E'(\mathbf{A}) = E(\mathbf{A}) - J_{ij}(a,b)\delta_{a_i,a}\delta_{a_j,b}
\end{equation}
The expectation value of any observable $O(\mathbf{A})$ under the pruned distribution $P' = e^{-E'}/Z'$ can be expressed in terms of averages over the original distribution $P = e^{-E}/Z$:
\begin{equation}
    \langle O(\mathbf{A}) \rangle_{P'} = \frac{\sum_{\mathbf{A}} O(\mathbf{A}) e^{-E'(\mathbf{A})}}{\sum_{\mathbf{A}} e^{-E'(\mathbf{A})}} = \frac{\langle O(\mathbf{A}) e^{J_{ij}(a,b)\delta_{a_i,a}\delta_{a_j,b}} \rangle_P}{\langle e^{J_{ij}(a,b)\delta_{a_i,a}\delta_{a_j,b}} \rangle_P}
\end{equation}
The symmetric Kullback-Leibler divergence between $P$ and $P'$ is thus given by:
\begin{equation}
\begin{split}
    D^{\mathrm{prune}}_{ij}(a,b) &= D_{KL}(P || P') + D_{KL}(P' || P) \\
    &= -\sum_{\mathbf{A}} [P(\mathbf{A}) - P'(\mathbf{A})][\ln P(\mathbf{A}) - \ln P'(\mathbf{A})] \\
    &= \langle E' - E \rangle_P - \langle E' - E \rangle_{P'} \\
    &= -J_{ij}(a,b) \langle \delta_{a_i,a}\delta_{a_j,b} \rangle_P + J_{ij}(a,b) \langle \delta_{a_i,a}\delta_{a_j,b} \rangle_{P'} \\
    &= J_{ij}(a,b) \left( p'_{ij}(a,b) - p_{ij}(a,b) \right)
\end{split}
\end{equation}
where $p_{ij}(a,b) = \langle \delta_{a_i,a}\delta_{a_j,b} \rangle_P$ is the marginal two-site probability under $P$, which coincides with the empirical frequency $f_{ij}(a,b)$ at convergence. Using the expression above, the marginal probability $p'_{ij}(a,b)$ under the pruned distribution $P'$ gives the following decimation score:
\begin{equation}
    D^{\mathrm{prune}}_{ij}(a,b) = J_{ij}(a,b) \left( \frac{p_{ij}(a,b)e^{J_{ij}(a,b)}}{p_{ij}(a,b)e^{J_{ij}(a,b)} + 1 - p_{ij}(a,b)} - p_{ij}(a,b) \right)
\end{equation}
The non-symmetrized KL divergence gives an equivalent ranking for this purpose; in practice, we remove the least significant couplings according to the lowest $D^{\mathrm{prune}}_{ij}(a,b)$ scores.

\subsubsection*{Training Algorithm} The edDCA training procedure alternates between pruning and tuning to maintain the target correlation as sparsity increases. Each epoch consists of the following steps:
\begin{itemize}
    \item \textbf{Pruning Step:} Identify and permanently remove a fixed fraction ($d_\mathrm{rate} = 1\%$) of the least significant active couplings from the graph $\mathcal{G}$.
    \item \textbf{Tuning Step (Sampling and Update):} If the Pearson correlation of the pruned model drops below the target threshold, we perform Monte Carlo sampling and gradient ascent updates on the remaining active parameters. This inner fine-tuning loop repeats until the correlation threshold is recovered. If the threshold is already met post-pruning, this step is skipped.
\end{itemize}
This alternating process is repeated until the target parameter density is reached while maintaining the required correlation threshold \cite{barrat2021sparse, rosset2024adabmdca}.

\subsection{Maximum Entropy Model Along Decimation Path (meDCA)}
\label{SI:meDCA}

The pruning and tuning steps described above have opposite effects on the entropy of the model distribution:

\begin{itemize}
    \item \textbf{Pruning parameters:} Removing parameters from the model (i.e., setting them to zero) relaxes the constraints on the sequence space. This shifts the model toward an independent-site profile model, which is less constrained. As a result, model entropy increases.
    \item \textbf{Tuning parameters:} Initializing from a flat uniform or profile model and tuning its parameters to fit a dataset strictly reduces its entropy. Although this reduction is not guaranteed when only the remaining parameters of an already trained model are updated, our Pearson correlation threshold of $\sim 0.95$ implies that the model is not fully converged to a perfect fit ($1.0$). Thus, tuning steps are expected to further reduce entropy.
\end{itemize}
The edDCA algorithm therefore involves a tradeoff between the entropy increase induced by pruning and the entropy reduction induced by tuning. Maximizing model entropy therefore requires removing only the least significant couplings, such as those dominated by finite-sampling noise in the data or in the training procedure, while minimizing updates of the remaining parameters.
Persistent Contrastive Divergence (PCD) accelerates training by initializing the Monte Carlo chains from the previous epoch's state and performing only a small number $K$ of sweeps over the sequence length (e.g., $K =5$ or $10$), which is usually sufficient to estimate the gradient direction. However, the same number of sweeps may be insufficient to accurately estimate the two-point marginal probabilities $p_{ij}(a,b)$ needed to identify the least significant couplings. Noisy estimates can therefore lead to the pruning of relevant couplings rather than purely noisy ones. Because such couplings contribute to the likelihood, their removal requires one or more tuning steps to recover the target Pearson correlation threshold, instead of allowing the algorithm to skip tuning. As a result, the entropy gained during pruning can be lost, or even overcompensated, during the subsequent tuning step.

To prevent this, we implemented a ``slow'' decimation strategy. We increased the number of Monte Carlo sweeps to $K=100$, approximately one order of magnitude larger than usual, to improve the precision of coupling selection. Furthermore, we reduced the pruning rate from $1\%$ of the active couplings per step down to $0.1\%$. This slower schedule preferentially removes the least significant parameters and largely avoids updates of the remaining parameters during the initial phase of decimation. In this regime, pruning dominates over tuning, producing an initial increase in model entropy up to a maximum.

\begin{figure}
\centering
\includegraphics[width=\textwidth]{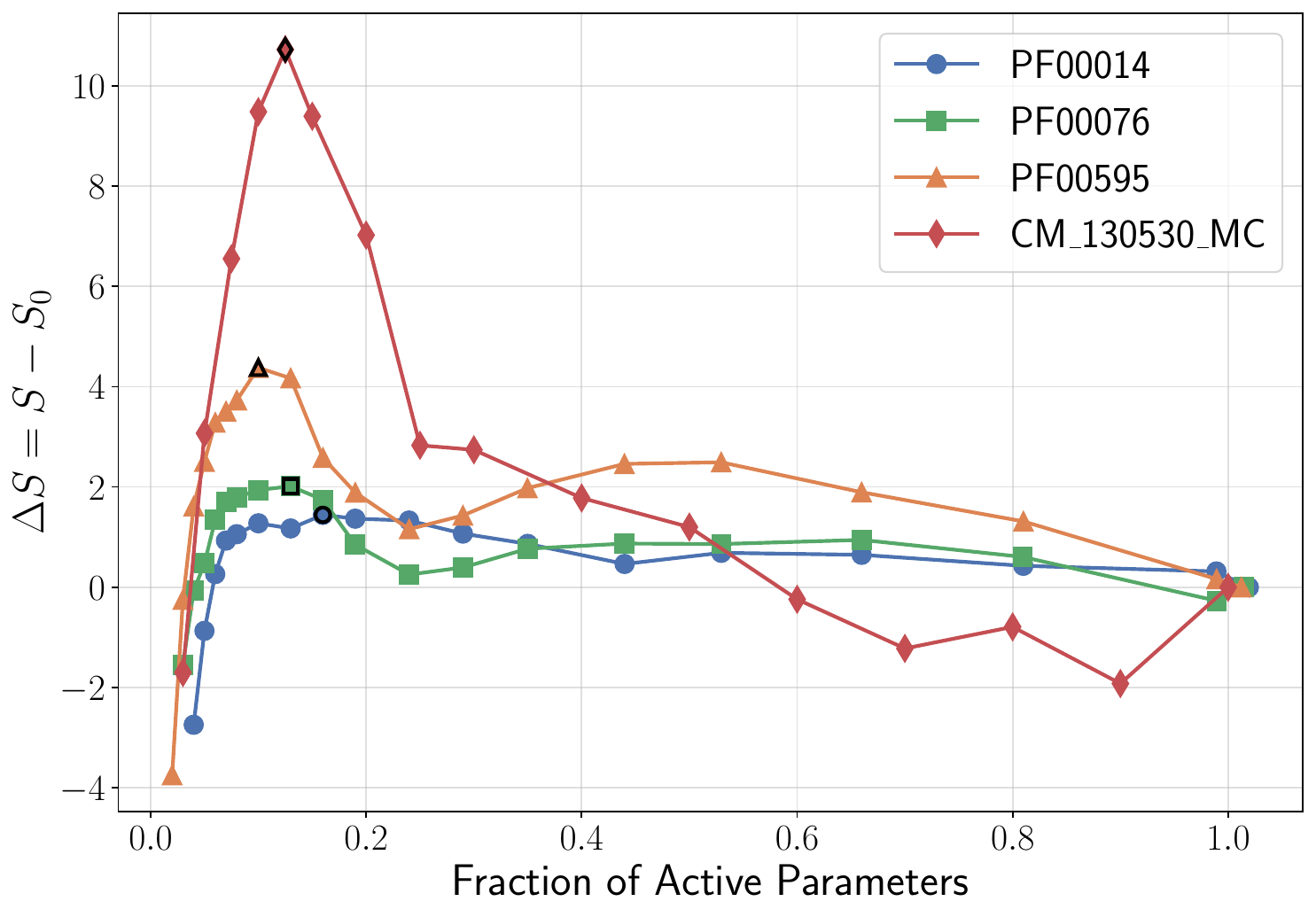}
\caption{\textbf{Entropy evolution during decimation across protein families.} The figure displays the variation in model entropy relative to the initial fully connected model ($\Delta S$) as a function of the fraction of active parameters for four diverse protein families. Despite differences in family size and sequence length, all datasets show a similar non-monotonic trajectory: an initial entropy increase driven by the removal of non-informative parameters, a maximum at intermediate density, and a rapid entropy decrease in the over-sparsified regime. This shared behavior confirms that the existence of an optimal high-entropy regime is a general feature of the decimation procedure and not specific to a single protein family.}
\label{SIfig:entropy_families}
\end{figure}

Below a critical parameter density, however, the balance shifts in the opposite direction. At this point, further pruning affects the likelihood, and the remaining parameters require large updates at each epoch to keep the Pearson correlation above the required threshold. Thus, although parameters are still being removed, the remaining explicitly fitted frequencies become more strongly constrained. In some cases, they must reach Pearson correlations close to $0.99$ just to keep the overall score above the selected threshold, leading to a substantial entropy reduction. As shown in the main text figures and in Fig.~\ref{SIfig:entropy_families}, this behavior defines an optimal parameter density corresponding to the entropy peak. Further sparsification below this threshold compromises the constraints needed to reproduce the amino acid statistics of the considered protein family.

\subsection{Edge Activation Algorithm (eaDCA)}
\label{SI:eaDCA}

The second strategy for training sparse models is based on Edge Activation (eaDCA). Unlike the top-down decimation procedure described above, eaDCA follows a bottom-up approach. It starts from an independent-site profile model, which captures only single-site conservation, and progressively activates parameters using an information-driven rule to increase the data likelihood.

\noindent The method was initially developed for RNA sequences \cite{calvanese2024parsimonious}, where parameters are activated block-wise, treating the entire $q \times q$ coupling matrix for a pair of positions as a single unit. For highly sparse protein models, this coarse-grained activation is suboptimal. We therefore adapt the analytical derivation to fine-grained, element-wise activation, where individual $(i,j,a,b)$ couplings are selected.

\subsubsection*{Element-wise Analytical Derivation:} 
Given a Multiple Sequence Alignment (MSA) of $M$ natural sequences, the log-likelihood of the Potts model at training step $t$ is expressed as:
\begin{equation}
\log(\mathcal{L}_t) = \sum_{r=1}^{M} \omega_r \log P_t(\mathbf{A^r}) = \sum_{r=1}^{M} \omega_r \left[ \sum_{(i,a) \in \mathcal{V}_t} h_i(a) \delta_{a_i^r, a} + \sum_{(i,j,a,b) \in \mathcal{E}_t} J_{i,j}(a,b) \delta_{a_i^r, a} \delta_{a_j^r, b} \right] - M_{\text{eff}} \log Z_t
\label{eq:log_likelihood}
\end{equation}
where $\mathbf{A^r}$ is the $r$-th sequence, $a_i^r$ is the amino acid at position $i$ in sequence $r$, $\omega_r$ is the corresponding weight ($M_{\text{eff}} = \sum_{r=1}^{M} \omega_r$) cf. \cite{cocco2018inverse} and $\mathcal{G}_t = (\mathcal{V}_t, \mathcal{E}_t)$ is the interaction graph at step $t$.

\noindent We consider the likelihood gain obtained by introducing or updating a single coupling parameter $\Delta J_{m,n}(a_m, a_n)$. The change in the log-likelihood function is given by:
\begin{equation}
\Delta \log(\mathcal{L}) = \log(\mathcal{L}_{t+1}) - \log(\mathcal{L}_t) = \Delta J_{m,n}(a_m, a_n) \sum_{r=1}^{M} \omega_r \delta_{A^r_m, a_m} \delta_{A^r_n, a_n} - M_{\text{eff}} \log \frac{Z_{t+1}}{Z_t}
\label{eq:delta_log_likelihood}
\end{equation}
The partition function ratio can be written as the expectation value of the perturbation under the current model distribution:
\begin{equation}
\frac{Z_{t+1}}{Z_t} = \frac{\sum_{\mathbf{a}} e^{-E_t(\mathbf{a}) + \Delta J_{m,n}(a_m, a_n)}}{\sum_{\mathbf{a}} e^{-E_t(\mathbf{a})}} = \langle e^{\Delta J_{m,n}(a_m, a_n)} \rangle_t
\label{eq:partition_ratio_expectation}
\end{equation}
Since $\Delta J_{m,n}$ contributes only when the amino acids $(a_m, a_n)$ occur at positions $(m,n)$, this expectation reduces to:
\begin{equation}
\frac{Z_{t+1}}{Z_t} = p_{m,n}(a_m, a_n) e^{\Delta J_{m,n}(a_m, a_n)} + 1 - p_{m,n}(a_m, a_n)
\label{eq:partition_ratio_simplified}
\end{equation}
where $p_{m,n}(a_m, a_n)$ is the marginal probability assigned by the current model to this specific co-occurrence.

\noindent The corresponding empirical frequency in the training dataset is:
\begin{equation}
f_{m,n}(a_m, a_n) = \frac{1}{M_{\text{eff}}} \sum_{r=1}^{M} \omega_r \delta_{A^r_m, a_m} \delta_{A^r_n, a_n}
\label{eq:empirical_frequency}
\end{equation}
Substituting these terms and normalizing by $M_{\text{eff}}$ gives the change in log-likelihood per effective sequence:
\begin{equation}
\frac{\Delta \log(\mathcal{L})}{M_{\text{eff}}} = \Delta J_{m,n}(a_m, a_n) f_{m,n}(a_m, a_n) - \log\left( p_{m,n}(a_m, a_n) e^{\Delta J_{m,n}(a_m, a_n)} + 1 - p_{m,n}(a_m, a_n) \right)
\label{eq:normalized_likelihood_change}
\end{equation}
Writing $p = p_{m,n}(a_m, a_n)$, $f = f_{m,n}(a_m, a_n)$, and $\Delta J = \Delta J_{m,n}(a_m, a_n)$, we maximize this expression by setting its derivative with respect to $\Delta J$ to zero:
\begin{equation}
0 = f - \frac{p e^{\Delta J}}{p e^{\Delta J} + 1 - p}
\label{eq:derivative_zero}
\end{equation}
Solving for $\Delta J$ yields:
\begin{equation}
\Delta J = \log \left( \frac{f(1-p)}{p(1-f)} \right)
\label{eq:optimal_delta_J}
\end{equation}
The second derivative is strictly negative, confirming that this stationary point is the global maximum. Substituting this optimal $\Delta J$ back into the normalized log-likelihood difference gives the maximum gain associated with this coupling:
\begin{equation}
\frac{\Delta \log(\mathcal{L})}{M_{\text{eff}}} = f \log \frac{f}{p} + (1-f) \log \frac{1-f}{1-p} = D_{KL}(f \parallel p)
\label{eq:max_likelihood_gain}
\end{equation}

\noindent Thus, at each step, the coupling $(m,n,a_m,a_n) \notin \mathcal{E}_t$ selected for activation is the one with the largest Kullback-Leibler divergence between the empirical frequency and the current model probability. Once selected, this interaction is activated, i.e. added to $\mathcal{E}_t$, and its value is updated by adding the analytically optimal $\Delta J$ derived above.

\noindent In the general $(i,j,a,b)$ notation, the activation step is defined as follows:
\begin{itemize}
    \item A coupling is selected for activation by maximizing the Kullback-Leibler divergence over all possible edges:
    \begin{equation}
        D^{\text{activate}}_{i,j}(a,b) = D_{KL}(f_{ij}(a,b) \parallel p_{ij}(a,b)) = f_{ij}(a,b) \log \frac{f_{ij}(a,b)}{p_{ij}(a,b)} + (1-f_{ij}(a,b)) \log \frac{1-f_{ij}(a,b)}{1-p_{ij}(a,b)}
        \label{eq:DKL_activate}
    \end{equation}
    \item Once selected, a coupling that is not already active is introduced into the model; if it is already active, its value is updated. In both cases, the parameter is adjusted by the analytically derived increment:
    \begin{equation}
        \Delta J_{ij}(a,b) = \log \left( \frac{f_{ij}(a,b)(1-p_{ij}(a,b))}{p_{ij}(a,b)(1-f_{ij}(a,b))}\right)
        \label{eq:eaDCA_update}
    \end{equation}
\end{itemize}

\subsubsection*{Training Algorithm:} Similar to the decimation approach, one epoch of the eaDCA training procedure is composed of distinct steps:
\begin{itemize}
    \item \textbf{Monte Carlo Sampling:} We perform $K$ Monte Carlo sweeps to estimate the two-point marginal probabilities $p_{ij}(a, b)$ of the current model, which are strictly required to evaluate Eq.~\eqref{eq:DKL_activate} and select the optimal coupling to update.
    \item \textbf{Activation Step:} This step analytically selects the parameter that yields the largest increase in data likelihood. The selected parameter might already belong to the active interaction graph $\mathcal{G}_t$, in which case it remains active and its value is updated. Because Eq.~\eqref{eq:eaDCA_update} gives the optimal update for that specific parameter, iterative gradient ascent is not required at this step, although gradient-based refinements can still be employed \cite{rosset2024adabmdca}.
\end{itemize}
This activation and updating procedure is repeated iteratively until the target Pearson correlation threshold is reached.

\noindent Since the model probabilities $p_{ij}(a,b)$ are estimated via Monte Carlo sampling, they are affected by sampling noise. Small statistical fluctuations can therefore inflate the Kullback-Leibler divergence of individual couplings. Without correction, the algorithm may select and activate couplings driven by noise rather than by reproducible statistical structure in the data. This instability is more pronounced in fine-grained, amino-acid-level activation models than in the original RNA implementation \cite{calvanese2024parsimonious}, and can hinder convergence.
To suppress the effect of sampling noise during coupling selection, we implemented a hierarchical, two-step selection process. First, we compute a block-wise score to select the pair of positions $(i,j)$ exhibiting the largest overall divergence. Only after the position pair is selected we evaluate the element-wise divergence to activate or update the specific amino acid coupling $(a,b)$ with the highest $D_{KL}$ within that block. 

\noindent The standard pseudocount $\alpha$, typically used in fully connected bmDCA to regularize the empirical one- and two-point frequencies $f_i(a)$ and $f_{ij}(a,b)$ and prevent diverging gradients, also provides an intrinsic parameter regularization here when applied directly in Eq.~\eqref{eq:eaDCA_update}, preventing singular parameter updates.

\section{Entropy Computation}
\label{SI:sample_entropy}

\subsection{Thermodynamic Integration} 
Estimating the Shannon entropy of a Potts model is computationally challenging because it requires evaluating the partition function $Z$, which involves a sum over the exponentially large sequence space of size $q^L$. We therefore compute the model entropy using Thermodynamic Integration (TI) \cite{marchi2019entropy}. The entropy $S$ is related to the average energy $\langle E \rangle$ and free energy $F$ through:
\begin{equation}
    S = \langle E \rangle_0 - F(0)
\end{equation}
where $\langle E \rangle_0$ is the average statistical energy of the sequences sampled from our unbiased DCA model, and $F(0) = -\ln Z(0)$ is its free energy. 
\noindent To calculate $F(0)$, TI constructs a continuous integration path between our target model (the unbiased DCA model, corresponding to a bias parameter $\theta = 0$) and a strongly biased model ($\theta = \theta_{\max}$) whose free energy can be easily evaluated. 
\noindent We define the biased model by introducing an external field $\theta$ that rewards sequences based on their similarity to a chosen target sequence $\textbf{A}_{wt}$ from the training set. The biased energy function takes the form:
\begin{equation}
    E_\theta(\textbf{A}) = E_0(\textbf{A}) - \theta \cdot \mathrm{SeqID}(\textbf{A}, \textbf{A}_{wt})
\end{equation}
where $E_0(\textbf{A})$ is the energy of the unbiased model, and $ \mathrm{SeqID}(\textbf{A}, \textbf{A}_{wt}) = \sum_{i=1}^L \delta_{a_i, a_{wt,i}}$ represents the sequence identity (the number of matching amino acids) between a generic sequence $\textbf{A}$ and the target sequence $\textbf{A}_{wt}$. 
\noindent The free energy of this biased model is $F(\theta) = -\ln Z(\theta)$. The derivative of the free energy with respect to the bias parameter $\theta$ is the negative average sequence identity:
\begin{equation}
    \frac{\partial F(\theta)}{\partial \theta} = -\frac{\partial}{\partial \theta} \ln \sum_\textbf{A} \exp\left(-E_0(\textbf{A}) + \theta \cdot \mathrm{SeqID}(\textbf{A}, \textbf{A}_{wt})\right) = - \langle  \mathrm{SeqID}(\textbf{A}, \textbf{A}_{wt}) \rangle_\theta
\end{equation}
where $\langle  \mathrm{SeqID}(\textbf{A}, \textbf{A}_{wt}) \rangle_\theta$ is the average sequence identity sampled from the distribution at a given bias $\theta$. 
\noindent By integrating this derivative from the unbiased state ($\theta = 0$) to a highly biased state ($\theta_{\max}$), we can express the free energy of our target model as:
\begin{equation}
    F(0) = F(\theta_{\max}) + \int_0^{\theta_{\max}} \langle  \mathrm{SeqID}(\textbf{A}, \textbf{A}_{wt}) \rangle_\theta \, d\theta
\end{equation}
To determine the boundary condition $F(\theta_{\max})$, we choose $\theta_{\max}$ large enough for the target sequence $\textbf{A}_{wt}$ to be sampled with measurable probability $p_{wt}$; in our implementation, $\theta_{\max}$ is chosen dynamically to obtain $p_{wt} \approx 0.1$. By definition, the probability of sampling the target sequence at maximum bias is given by $p_{wt} = \exp(-E_{\theta_{\max}}(\textbf{A}_{wt}) - F(\theta_{\max}))$. Rearranging this expression allows us to exactly compute the free energy at the boundary:
\begin{equation}
    F(\theta_{\max}) = \ln(p_{wt}) + E_{\theta_{\max}}(\textbf{A}_{wt})
\end{equation}
Substituting $F(0)$ back into the initial thermodynamic relation, we obtain the final formula used to compute the entropy of the DCA model:
\begin{equation}
    S = \langle E_0 \rangle_0 - \left[ \ln(p_{wt}) + E_{\theta_{\max}}(\textbf{A}_{wt}) + \int_{0}^{\theta_{\max}} \langle  \mathrm{SeqID}(\textbf{A}, \textbf{A}_{wt}) \rangle_\theta \, d\theta \right]
\end{equation}
In practice, the integral is computed numerically. We discretize the interval $[0, \theta_{\max}]$ linearly. At each integration point, Monte Carlo sampling under the biased energy $E_\theta(\textbf{A})$ is used to estimate $\langle  \mathrm{SeqID}(\textbf{A}, \textbf{A}_{wt}) \rangle_\theta$, and the integral is then evaluated using the trapezoidal rule \cite{marchi2019entropy}.

\subsection{Subspace Entropy and Effective Support Size}

Our objective is to estimate the model's entropy restricted to specific sub-regions of the sequence space. We define these sub-regions based on the sequence divergence from the closest natural wild-type (WT) sequence in the training alignment. 
As detailed in the previous section, Thermodynamic Integration (TI) provides a method to compute the free energy $F$ of the Potts model. Knowing the free energy allows us to directly extract the log-partition function, $\ln Z = -F$. Given $\ln Z$, the statistical probability $P(\textbf{A})$ assigned by the model to any generated sequence $\textbf{A}$ is:
\begin{equation}
    \ln P(\textbf{A}) = -E(\textbf{A}) - \ln Z
\end{equation}
For a subspace defined by a sequence-divergence interval $D$, e.g. 20--25\% from the closest WT, we estimate the entropy of the model distribution restricted to that region. This local entropy can be approximated from a sufficiently large set of independently generated model samples. Specifically, the local entropy $S_D$ is estimated as the sample mean of the negative log-probabilities of the sequences that fall into the divergence interval $D$:
\begin{equation}
    S_D \approx - \frac{1}{N_D} \sum_{\textbf{A} \in D} \ln P(\textbf{A}) = \langle E(\textbf{A}) \rangle_D + \ln Z
\end{equation}
where $N_D$ is the number of sampled sequences in the interval, and $\langle E(\textbf{A}) \rangle_D$ is their average statistical energy. 
To obtain sufficient statistics, we first generated a large unconditional sequence sample from the model using standard Monte Carlo methods. This sample was then filtered into the sequence-divergence intervals analyzed in the main text.

\noindent A methodological limitation is that the entropy could not be reliably estimated in the very low sequence-divergence range ($0$--$20\%$). Generative models of this class do not spontaneously sample enough sequences so close to the training data, making the estimate above unreliable in this range.

\noindent Finally, once the local entropy $S_D$ for a specific divergence interval is computed, we can evaluate the effective support size of the model in that region. This quantity, denoted $\Omega_D$, represents the effective number of sequences generated by the model within that distance from natural sequences \cite{lambert2024expanding}:
\begin{equation}
    \Omega_D = e^{S_D}
\end{equation}
This quantity provides a measure of the effective sequence-space volume explored by different models at fixed distances from natural sequences.

\section{Experimental Procedure}
\label{SI:experiment}

\subsection{Chorismate mutase library construction}

Genes encoding the designed protein sequences were ordered as a mixed pool of single-stranded oligonucleotides (IDT oPools). Nucleotide sequences corresponding to the designed amino acid sequences were determined by randomly sampling codons proportionally to their usage in K12 \textit{Escherichia coli}~\cite{nakamura2000codon}. Two stop codons (TAATGA) were added to the end of the coding sequence to minimize read-through. The 5' end of the coding sequence was preceded by a primer site for GSP2 (5’-AAACCGGAGCCATACAGTAC-3’) and an NdeI restriction site (5'-CATATG-3'), and the stop codons at the 3' end of the coding sequence are followed by an XhoI restriction site (5'-CTCGAG-3') and a reverse complement priming site for GSP107 (5'-CCGTGCGACAAGATTTCAAG-3’)~\cite{subramanian2018set}. Oligonucleotides ranged from 300bp to 346bp in length. If a design yielded an oligonucleotide fewer than 300bp, the sequence was padded with random nucleotides between the stop codons and the XhoI restriction site to a final length of 300bp. Final oligonucleotides were free of internal priming sites for GSP2 or GSP107 (defined as a 10 bp or more match on either strand), internal restriction sites, or internal repeated subsequences of 10 bp or more. No more than 9 repeated A or repeated T bases and no more than 5 repeated G or repeated C bases were allowed. The GC content of both the coding sequence and the full oligonucleotide was restricted to be between 40\% and 60\%. The GC content of the first 60 (5') bases was restricted to be less than 65\%. No 100 bp window of the oligonucleotide had a GC content above 70\%, no 20 bp window had a GC content above 90\%, and the highest and lowest GC content within any 50 bp window differed by no more than 50\%.
\noindent Oligonucleotide pools were amplified by PCR using KAPA HiFi DNA polymerase (Roche) with 1$\times$ KAPA HiFi Fidelity buffer, 0.2 mM dNTPs 2 mM GSP2, 2 mM GSP107, 0.02 $\mu$L KAPA HiFi DNA polymerase, and 0.4 ng/$\mu$L template DNA in a final reaction volume of 50 $\mu$L. The reaction was incubated at 95$^\circ$C for 3 min (polymerase activation), followed by 14 cycles of 98$^\circ$C for 20 s (denaturation), 60$^\circ$C for 15 s (annealing), and 72$^\circ$C for 15s (extension). After 14 cycles, a final extension was performed by incubating the reaction at 72$^\circ$C for 2 min. To separate the desired PCR products from DNA synthesis artifacts, amplicons of the target size (300-346 bp) were gel purified and concentrated (Zymo Research). Purified amplicons were digested with NdeI and XhoI (NEB), ligated into NdeI- and XhoI-digested pKTCTET-0~\cite{roderer2014functional}, column purified (Zymo Research), and transformed into XL1-Blue \textit{E. coli} (Agilent) to yield $>$1000X coverage per gene. The entire transformation was cultured in 100 mL LB media containing 100 $\mu$g/mL sodium ampicillin at 37$^\circ$C overnight, after which plasmids were purified via miniprep (Macherey Nagel).

\subsection{Chorismate mutase selection assay}

A previously reported protocol was used to carry out the selection assay~\cite{russ2020evolution, roderer2014functional}. It should be noted that the meDCA sequences were evaluated in an independent experimental batch from the other three models, but the inclusion of natural sequences as internal controls in both assays enables robust comparability, as evidenced by the strong correlation of relative enrichment scores across the two experiments. The library of Potts model designs in pKTCTET-0, a previously constructed library of natural CM coding sequences in pKTCTET-0~\cite{russ2020evolution}, and a previously constructed standard curve library of \textit{E. coli} CM point mutants in pKTCTET-0~\cite{russ2020evolution} were each diluted to a concentration of 0.5 ng/mL and separately transformed into the CM-deficient assay strain KA12/pKIMP-UAUC~\cite{kast1996exploring} to yield $>$1000X coverage per gene. Transformation mixtures were incubated for one hour at 37$^\circ$C immediately following transformation, then diluted separately into 500 mL LB media containing 100 $\mu$g/mL sodium ampicillin and 30 $\mu$g/mL chloramphenicol and incubated overnight at 30$^\circ$C. Aliquots of each overnight culture maintaining $>$1000X coverage of each gene were supplemented with 20\% glycerol and frozen at -80$^\circ$C.

\noindent Glycerol stocks of KA12/pKIMP-UAUC carrying each library were revived separately in LB media containing 100 $\mu$g/mL sodium ampicillin and 30 $\mu$g/mL chloramphenicol overnight at 30$^\circ$C. Overnight cultures were used to inoculate separate cultures of M9c minimal media~\cite{roderer2014functional} supplemented with 100 $\mu$g/mL sodium ampicillin, 30 $\mu$g/mL chloramphenicol, 20 $\mu$g/mL of L-phenylalanine (F), and 20 $\mu$g/mL of L-tyrosine (Y) (M9cFY, non-selective conditions) to an OD$_{600}$ of 0.045. M9cFY cultures were grown to 30$^\circ$C to an OD$_{600}$ of approximately 0.2. Cultures were mixed to yield equal representations of each gene in the designed and natural libraries, and 4X representation of each gene in the standard curve. The mixed culture was washed twice with M9c media (no FY). The washed culture was used to inoculate 500 mL M9c (selection) and 500 mL M9cFY (non-selection control), each supplemented with 100 $\mu$g/mL sodium ampicillin, 30 $\mu$g/mL chloramphenicol, and 3 ng/mL doxycycline hyclate (to induce gene expression) to an OD$_{600}$ of 0.001. The remainder of the mixed culture was harvested and resuspended in 2 mL LB with 100 $\mu$g/mL sodium ampicillin and 30 $\mu$g/mL chloramphenicol, grown overnight at 37$^\circ$C, and harvested for plasmid purification (the input sample) via miniprep (Macherey Nagel). The selection and non-selection cultures were grown for 24h at 30$^\circ$C, passaging as necessary to maintain an OD$_{600} < $ 0.1. For both the selection and non-selection cultures, 50 mL of culture was harvested, resuspended in 2 mL LB with 100 $\mu$g/mL sodium ampicillin and 30 $\mu$g/mL chloramphenicol, grown overnight at 37$^\circ$C, and harvested for plasmid purification via miniprep.

\noindent Input, selected, and non-selected samples were amplified using two rounds of PCR with Q5 High-Fidelity DNA polymerase (NEB) to add adapters and indices for Illumina sequencing. The first round of PCR (2 cycles) added 6-9 random nucleotides (for initial focusing during sequencing) and part of the i5 and i7 Illumina adapters. The products of the first round of PCR were purified via bead-cleanup (Beckman Coulter). The second round (18 cycles) added the remaining i5 and i7 adapters as well as TruSeq indices to each sample. In addition to a low number of total cycles, a high initial DNA concentration in both rounds was used to minimize amplification bias. Second round PCR products were gel purified (Zymo Research), quantified via Qubit (ThermoFisher), mixed in equimolar ratios, diluted to 4 nM, and sequenced on an Illumina MiSeq system with a 2x300 paired-end kit.

\subsection{Experimental data processing}

Paired-end Illumina reads were joined using FLASH~\cite{magoc2011flash}, trimmed to the NdeI and XhoI restriction sites, and translated. For the designed and natural libraries, only exact matches to library variants were counted. For the standard curve point mutants, only exact match sequences with a correct 4 bp barcode were counted.  Any sequence with fewer than 11 reads in the input sample was excluded from analysis. After filtering for low initial count, a pseudocount of 1 read per sequence per sample was added for calculations. Relative enrichment scores (r.e.) for the selected sample were calculated as

$$r.e. = \log \frac{f_s^x}{f_i^x} - \log \frac{f_s^r}{f_i^r}.$$

\noindent The variables $f_s^x$ and $f_i^x$ are the frequencies of each allele \textit{x} in the selected (\textit{s}) or input (\textit{i}) samples. The variables $f_s^r$ and $f_i^r$ are the corresponding values for the \textit{E. coli} chorismate mutase, the reference allele. For comparison between experiments and to published data, normalized relative enrichment scores were calculated such that a value of 1 corresponds to the \textit{E. coli} wild-type and a value of 0 corresponds to the mean of the near-null mode across all libraries as fit by a two-mode Gaussian mixture model. Functional labels were assigned to variants with relative enrichment scores falling more than 3 standard deviations above the mean of the near-null mode.

\section{Context-Dependent Entropy and Mutability}
\label{SI:CDE}

To quantify the mutability of a specific protein sequence, we rely on entropy as an information-theoretic measure of mutational permissiveness. Although mutability can be assessed globally across a protein family, here we focus on how a \textit{specific sequence context} constrains the allowed mutational space. To do this, we distinguish between two definitions of site-specific entropy: Context-Independent Entropy (CIE) and Context-Dependent Entropy (CDE) \cite{figliuzzi2016coevolutionary}. Averaging the CDE across all sites then gives an overall measure of sequence mutability.

\subsubsection*{Context-Independent Entropy (CIE)}
The CIE provides a measure of global mutability for a given site $i$, as observed across all natural sequences in a family. It is estimated directly from the empirical frequencies $f_i^\mathrm{nat}(a)$ of amino acid $a$ at column $i$ of the multiple sequence alignment (MSA) of natural sequences:
\begin{equation}
\mathrm{CIE}_i = -\sum_a f_i^\mathrm{nat}(a) \log f_i^\mathrm{nat}(a).
\end{equation}
Because our generative model aims to reproduce the natural data distribution, the empirical frequency $f_i^\mathrm{nat}(a)$ closely approximates the model's marginal probability $p_i(a)$. Therefore, the CIE captures the mutability of a site in evolutionary equilibrium, entirely independent of any specific sequence context.

\subsubsection*{Context-Dependent Entropy (CDE) and Sequence Mutability}
In contrast to the CIE, the Context-Dependent Entropy (CDE) expresses the mutability of a site within a specific amino acid context. To compute the CDE, we evaluate the conditional probability $P(a_i = a | \mathbf{A}_{\backslash i})$ assigned by the model for all possible amino acids at site $i$, given the fixed sequence background $\mathbf{A}_{\backslash i} = (a_1, ..., a_{i-1}, a_{i+1}, ..., a_L)$. The CDE is calculated as:
\begin{equation}
\mathrm{CDE}_i(\mathbf{A}_{\backslash i}) = -\sum_a P(a_i = a | \mathbf{A}_{\backslash i}) \log P(a_i = a | \mathbf{A}_{\backslash i}).
\end{equation}
Unlike the CIE, the CDE explicitly depends on the chosen sequence background and requires an inferred statistical generative model to be evaluated. Because the CDE depends on the background sequence, its value changes with the sequence context, for example when evaluating a distant homolog.

\noindent We define the overall \textit{mutability} of a protein sequence $\mathbf{A}$ as the average CDE across its $L$ sites. This average provides a global, context-aware measure of mutational permissiveness. A higher average CDE indicates that the sequence lies in a broader local neutral space, where more mutations are compatible with the statistical constraints of the protein family.

\newpage

\begin{figure}
\centering
\includegraphics[width=\textwidth]{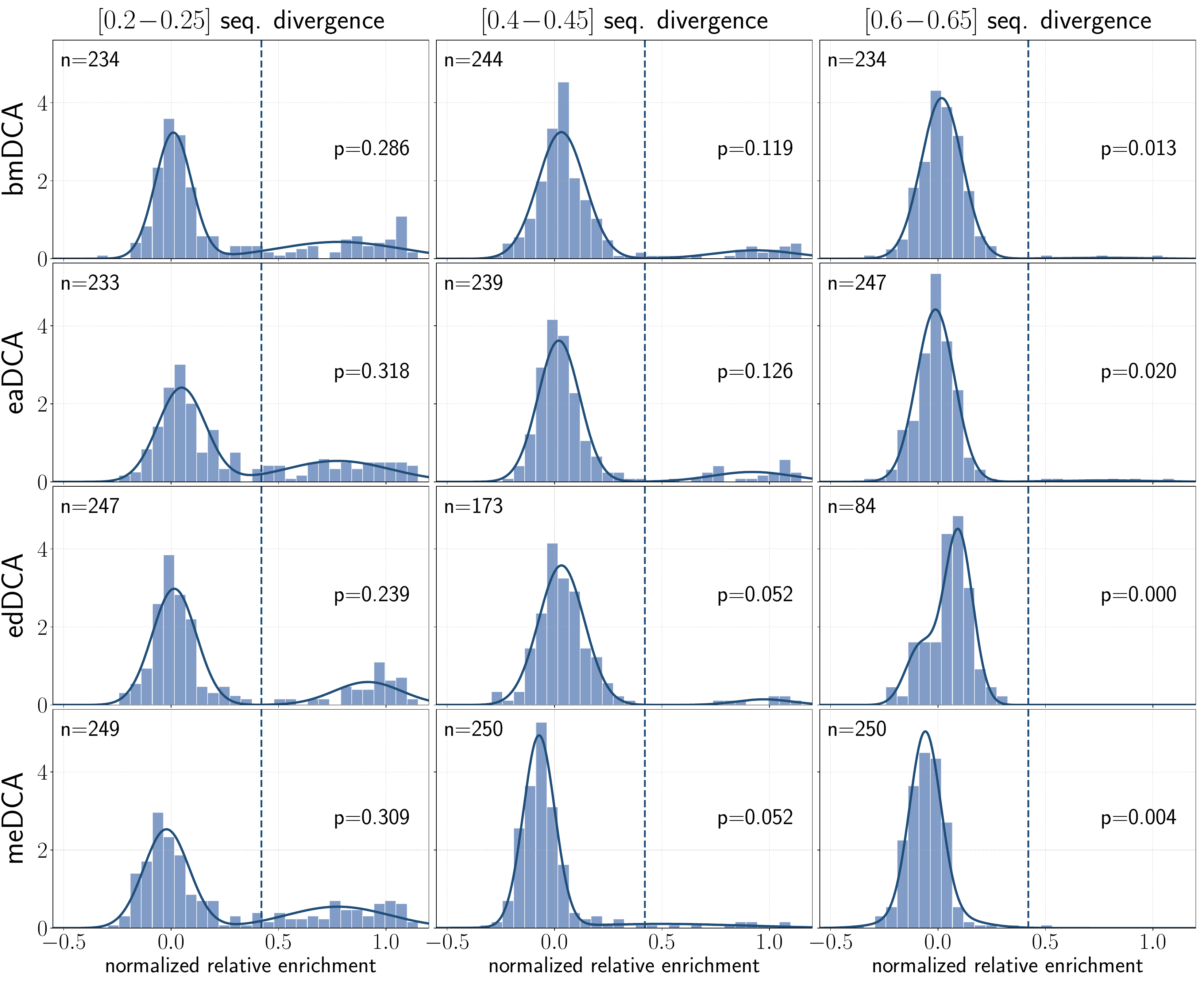}
\caption{\textbf{Experimental functional evaluation of artificial sequences across models and divergence intervals.} 
    The 4$\times$3 grid displays histograms of the normalized relative enrichment scores obtained from the \textit{in vivo} complementation assay. 
    Rows correspond to the four evaluated generative models (bmDCA, eaDCA, edDCA, and meDCA). 
    Columns represent the three targeted ranges of sequence divergence from the closest natural homolog in the training set (20--25\%, 40--45\%, and 60--65\%). 
    Solid lines show the fitted bimodal Gaussian mixture model used to separate the near-null mode, corresponding to non-functional sequences, from the functional sequence distribution.
    In each panel, $n$ denotes the number of artificial sequences tested for the corresponding model and divergence interval, and $p$ denotes the fraction classified as functional.}
\label{SIfig:experiment}
\end{figure}

\begin{figure}
\centering
\includegraphics[width=1.\textwidth]{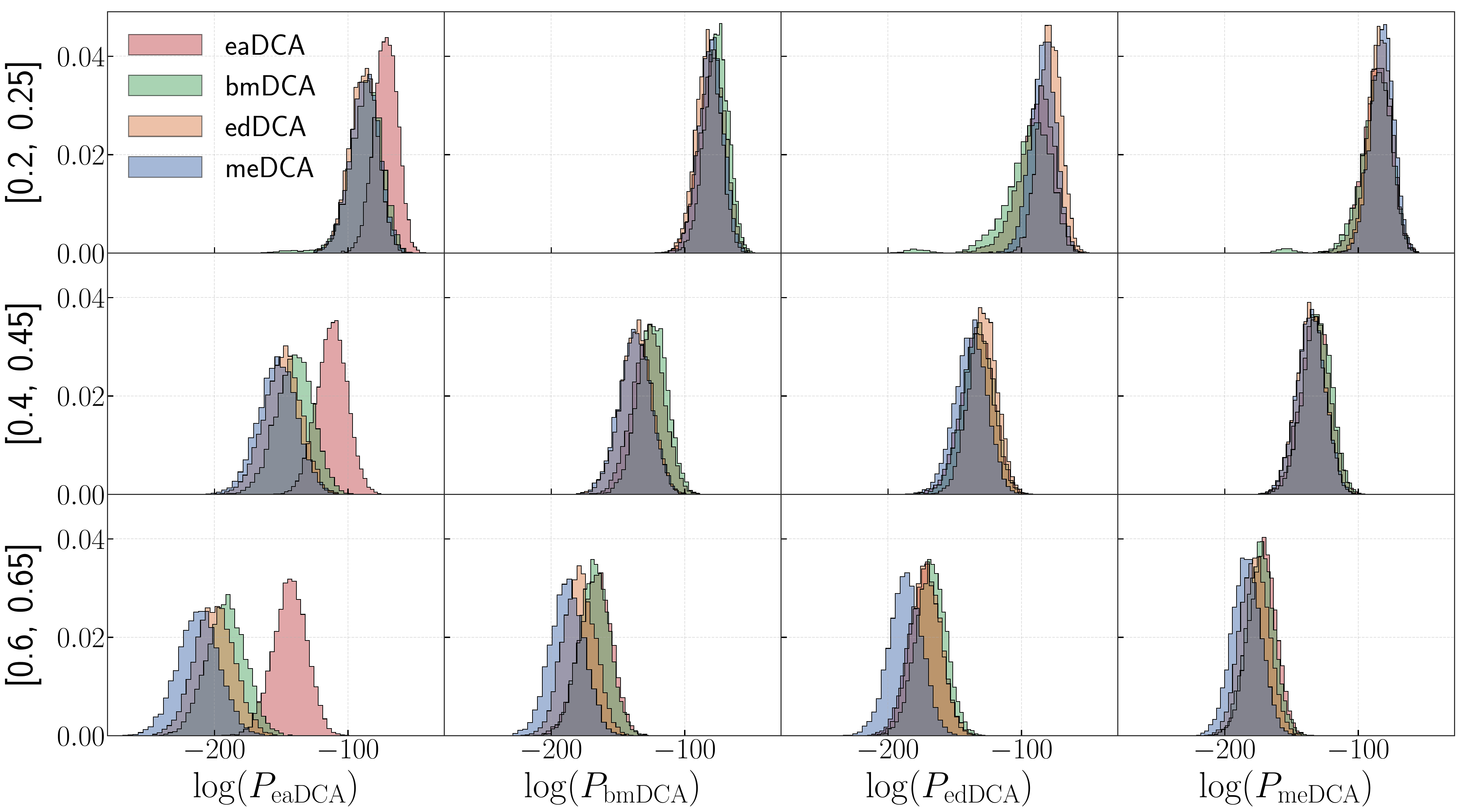}
    \caption{\textbf{Comprehensive cross-model scoring analysis across all sequence divergence intervals.} 
    The 3$\times$4 grid displays histograms of the statistical energy (log-probability) assigned by a specific ``scoring'' model (columns) to synthetic sequences generated by all four models (indicated by different colors). 
    The three rows correspond to the targeted sequence divergence intervals from the closest natural homolog in the training set: close (20--25\%), moderate (40--45\%), and distant (60--65\%). 
    The four columns represent the specific model used to score the sequences (eaDCA, bmDCA, edDCA, and meDCA, respectively). 
    Expanding upon the analysis in the main text, these distributions confirm a systematic asymmetry across all divergence levels: high-entropy models (such as meDCA) assign high probabilities to sequences generated by lower-entropy models, recognizing them as part of their viable sequence space. Conversely, low-entropy models (such as eaDCA) assign systematically lower probabilities to sequences generated by broader, higher-entropy models, highlighting their narrower representation of the fitness landscape.}
\label{SIfig:cross-scoring}
\end{figure}

\begin{figure}
    \centering
    \includegraphics[width=\linewidth]{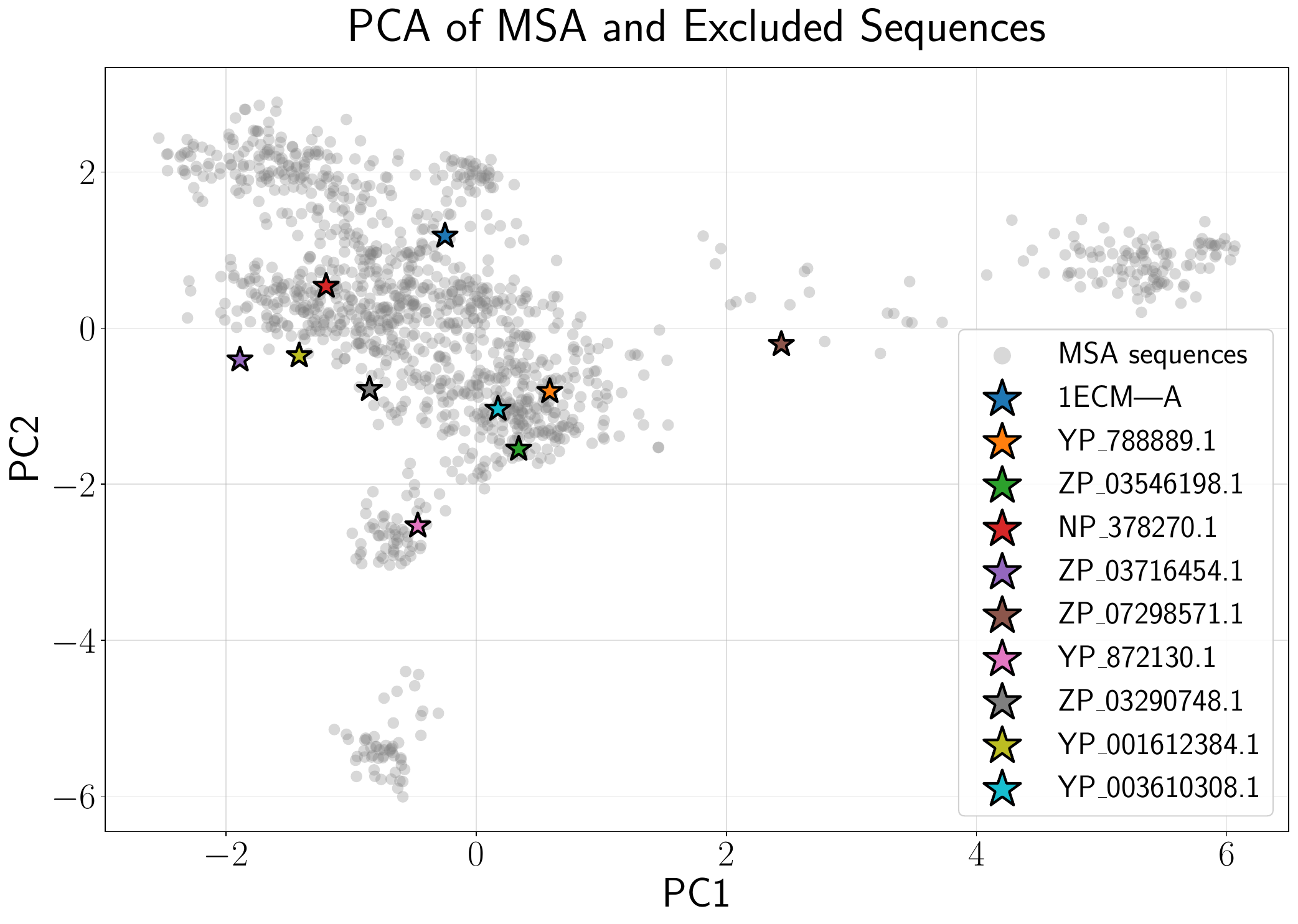}

        \caption{\textbf{PCA projection of the training MSA and excluded sequences for leave-one-group-out validation.} 
    The scatter plot displays the first two principal components (PC1 vs. PC2) of the natural wild-type sequences comprising the Chorismate Mutase training alignment (background points). Highlighted on top of this distribution are the 10 specific natural sequences selected as targets for the generalization and overfitting analysis discussed in the main text. To test the models' generative capacity in unobserved regions of sequence space, each of these 10 targets, together with its local neighborhood, defined as all sequences with $<20\%$ sequence divergence from the target, was removed from the alignment to train distinct subset models. This projection shows that the excluded sequences are well separated and broadly distributed across the natural sequence space, providing a diverse set of held-out targets for evaluating generalization beyond the training data.}
    \label{SIfig:PCAexcluded}
\end{figure}

\begin{figure}
    \centering
    \includegraphics[width=\linewidth]{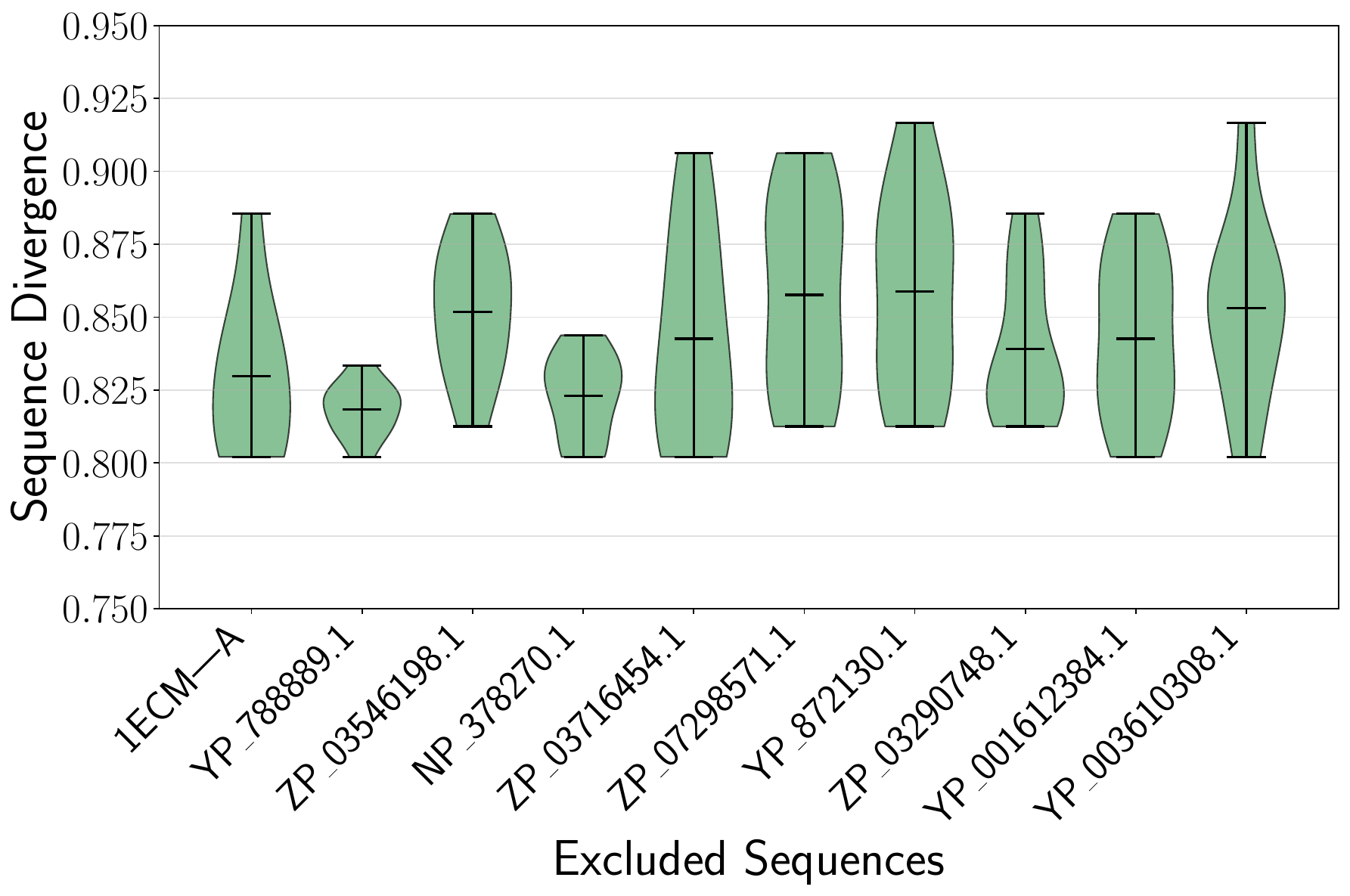}
    \caption{\textbf{Sequence divergence distribution of the excluded target sequences.} 
    For each of the 10 natural target sequences, the violin plots report the distribution of sequence divergences to the remaining natural sequences in the alignment. These targets were used to define the held-out groups in the leave-one-group-out validation analysis described in the main text. The observed divergence profiles show that the selected targets span distinct regions of natural sequence space, providing a stringent test of model generalization and overfitting.}
    \label{SIfig:seq_div_excluded}
\end{figure}

\begin{figure}
        \centering
        \includegraphics[width=1\linewidth]{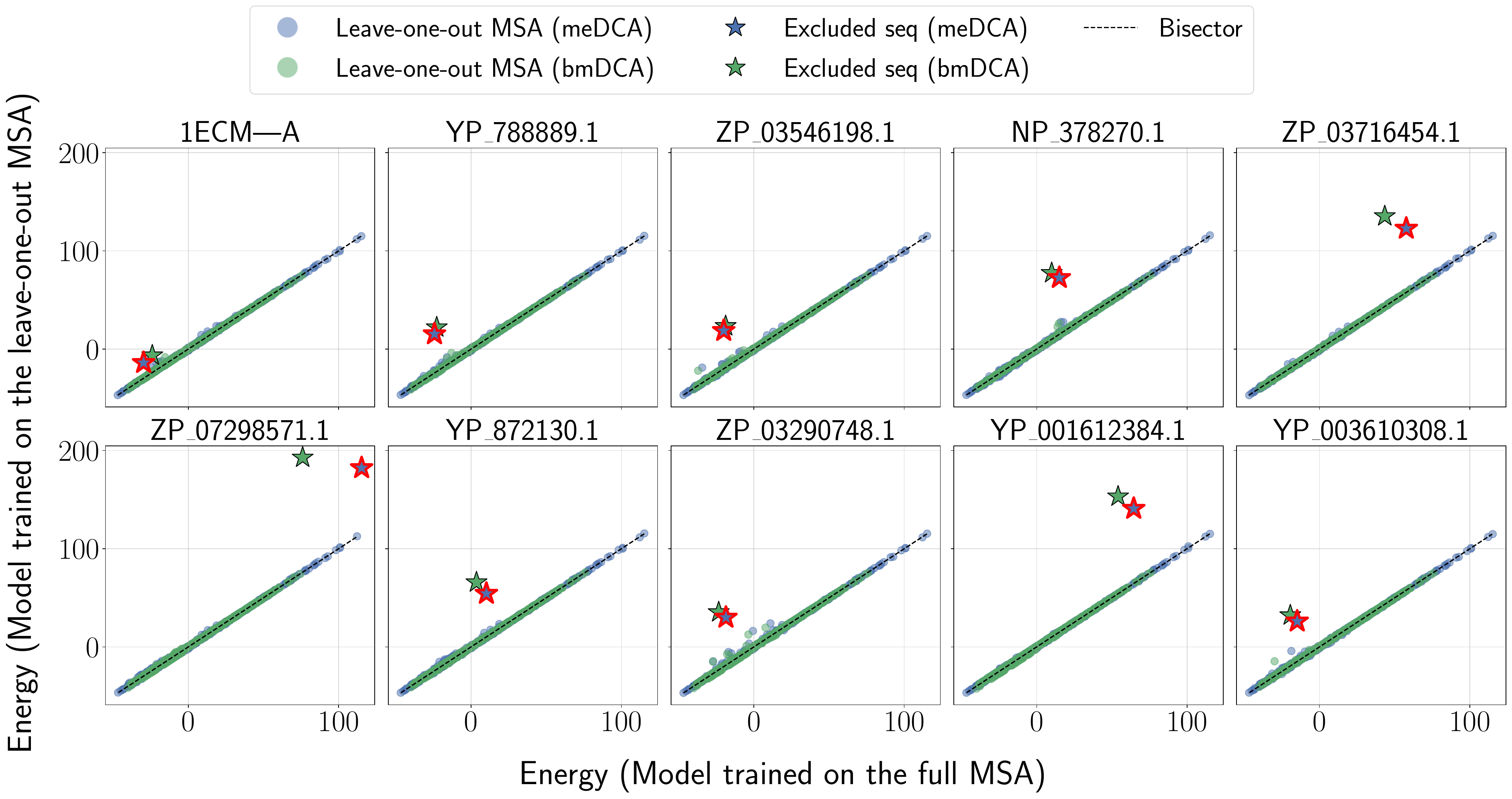}
        \caption{\textbf{Effect of leave-one-group-out retraining on the statistical energy.}
        For each of the ten natural Chorismate Mutase sequences selected for the leave-one-group-out validation test, we compare the statistical energies assigned by bmDCA and meDCA models trained on the full MSA with those assigned by the corresponding models retrained after removing the target sequence and its close homologs. Each panel corresponds to one excluded sequence. Background points show the energies of the sequences in the corresponding leave-one-out MSA, with bmDCA shown in green and meDCA in blue, while highlighted stars indicate the excluded target sequence scored by the two models. The diagonal line marks equal energy under the full-MSA and leave-one-out models. Removing a local neighborhood from the training set has almost no effect on the energies assigned to the remaining natural sequences; in contrast, compared with bmDCA, meDCA systematically assigns lower energies to the withheld target sequences, highlighted with a red outline. This indicates that meDCA is less ``surprised'' by previously unseen natural sequences, assigning them energies closer to those of the training-set sequences. This analysis provides a complementary view of the overfitting assessment reported in the main text.}
    \label{fig:S8}
\end{figure}

\begin{figure}
\centering
\includegraphics[width=1\textwidth]{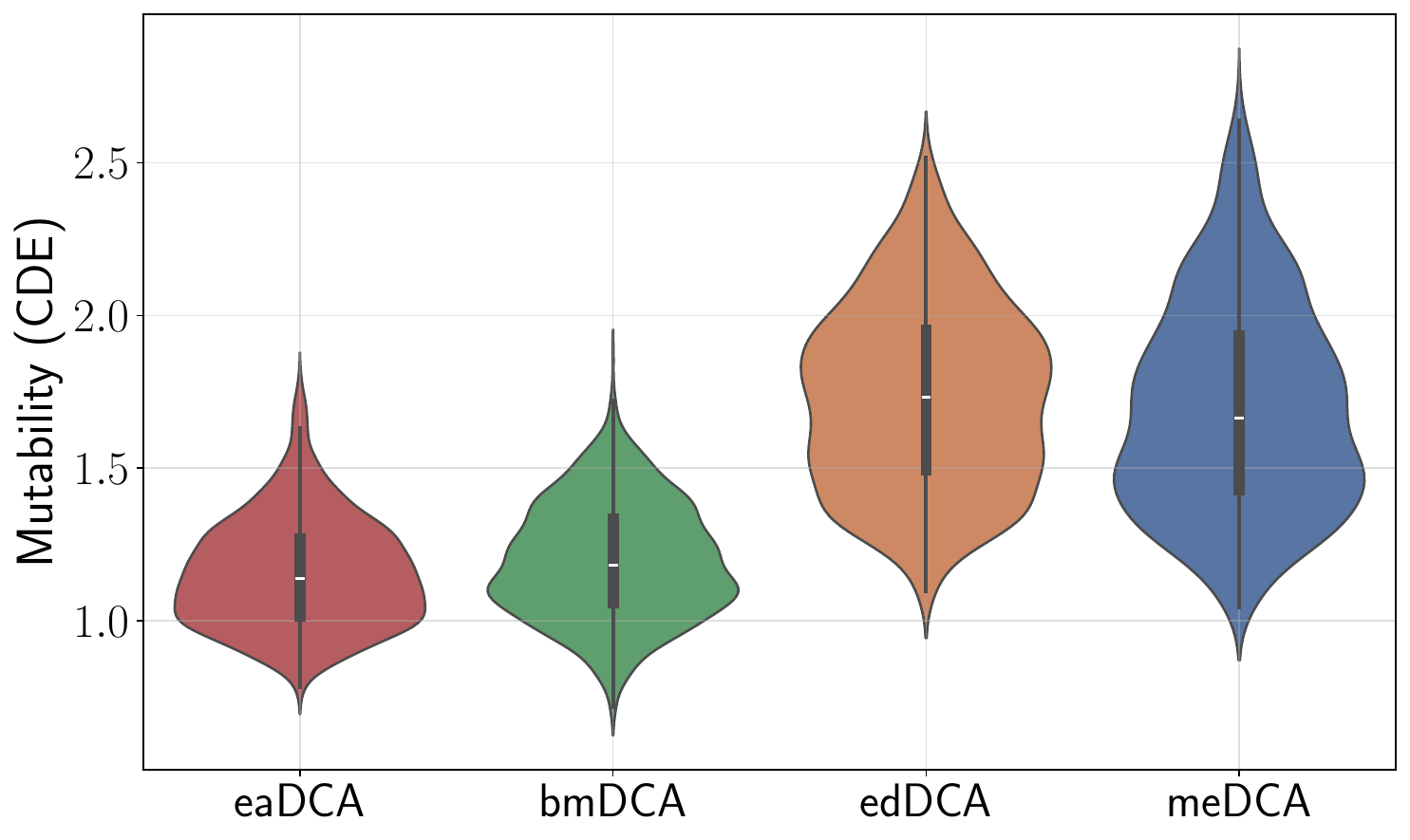}
\caption{\textbf{Mutability of natural sequences across all generative models.} 
    The violin plots display the distribution of mutability scores for natural Chorismate Mutase sequences from the training alignment. Mutability is quantified as the site-specific Context-Dependent Entropy (CDE) averaged over the length of each sequence. The four evaluated models are ordered on the x-axis from left to right by increasing model entropy (eaDCA, bmDCA, edDCA, meDCA). Expanding upon the analysis in the main text, a clear trend emerges: there appears to be a distinct threshold in model entropy beyond which the mutability of natural sequences significantly increases. Low-entropy models impose stronger constraints on natural sequences, resulting in lower mutability scores. In contrast, higher-entropy models assign higher mutability, consistent with a broader local neutral space around wild-type sequences. This suggests that sufficiently high model entropy is needed to capture the local mutational permissiveness of the fitness landscape.}
\label{SIfig:CDE}
\end{figure}

\begin{figure}
    \centering
    \includegraphics[width=1\linewidth]{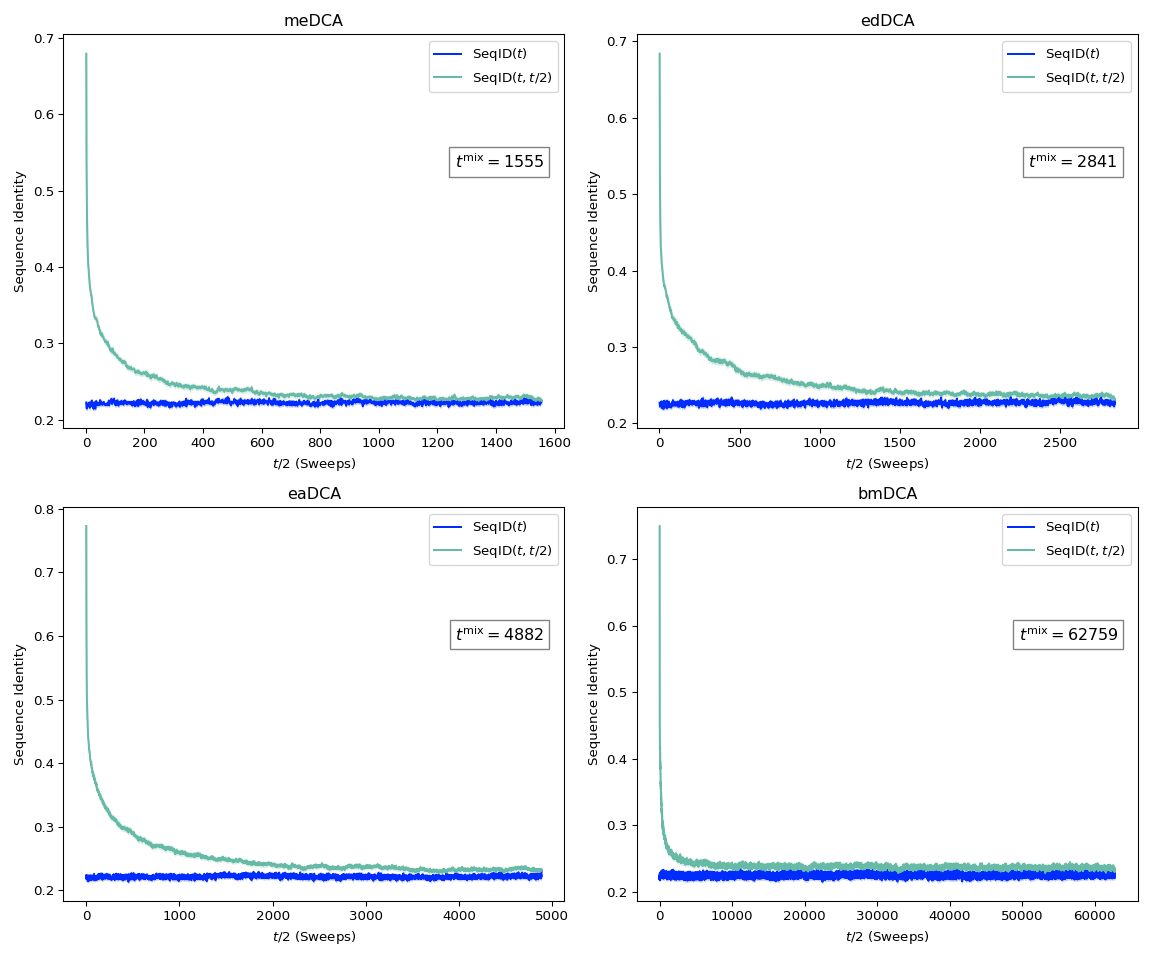}
    \caption{\textbf{Comparison of mixing times across the four evaluated models.} The four subplots display the sampling dynamics for meDCA, edDCA, eaDCA, and bmDCA, respectively. To quantify the mixing time, the average sequence overlap between a sample of Markov chains at time $t$ and time $t/2$ is plotted as a function of $t/2$. The reference baseline in each subplot represents the expected sequence overlap of two completely independent samples, computed by randomly permuting the sample with itself. The mixing time of each model is defined as the point at which the $t$ vs. $t/2$ overlap decays and converges to this independent baseline. As expected, the highest-entropy model (meDCA) exhibits a significantly faster mixing time compared to the others. Specifically, meDCA mixes approximately 2 times faster than edDCA, 3 times faster than eaDCA, and about 40 times faster than bmDCA.}
    \label{SIfig:mixingtime}
\end{figure}

\clearpage

\begin{table}[ht]
\centering

\caption{\textbf{Comparison of GPT architectures under two stopping protocols.}
For each GPT architecture, we report the number of trainable parameters, the selected stopping epoch, the Pearson correlation between model-generated and natural-data pairwise correlations $C_{ij}$, the estimated entropy, the final training loss, and the learning rate. Architecture names encode the embedding dimension (E), number of attention heads (H), and number of Transformer layers (L); for example, GPT-E64-H4-L4 denotes a model with embedding dimension 64, 4 attention heads, and 4 layers. Two stopping protocols are compared: \emph{Validation-loss stop}, where a validation split is used to identify the epoch of minimum validation loss and the model is then retrained on the full dataset up to that epoch; and \emph{$C_{ij}$ Pearson stop}, where training is stopped when the Pearson correlation between model-generated and natural $C_{ij}$ exceeds $0.94$. For the smallest architecture, GPT-E64-H4-L4, this threshold was never reached; therefore, we selected the first epoch at which the model reached its maximum attainable $C_{ij}$ Pearson correlation, approximately $0.8$.}
\label{tab:gpt_stopping}
\begin{tabular}{llrrrrrr}
\hline
\textbf{GPT architecture} &
\textbf{Stopping protocol} &
\textbf{\# params} &
\textbf{Stop epoch} &
\textbf{$C_{ij}$ Pearson} &
\textbf{Entropy} &
\textbf{Train loss} &
\textbf{LR} \\
\hline
GPT-E64-H4-L4  & Validation-loss stop & 202,112   & 500   & 0.710 & 144.89 & 1.4573 & 0.001 \\
GPT-E64-H4-L4  & $C_{ij}$ Pearson stop & 202,112   & 4,000 & 0.806 & 87.43  & 0.8066 & 0.001 \\
\hline
GPT-E128-H4-L4 & Validation-loss stop & 797,440   & 320   & 0.663 & 152.70 & 1.4996 & 0.001 \\
GPT-E128-H4-L4 & $C_{ij}$ Pearson stop & 797,440   & 1,600 & 0.943 & 29.53  & 0.2894 & 0.001 \\
\hline
GPT-E128-H8-L6 & Validation-loss stop & 1,192,960 & 180   & 0.712 & 145.05 & 1.4224 & 0.001 \\
GPT-E128-H8-L6 & $C_{ij}$ Pearson stop & 1,192,960 & 900   & 0.942 & 24.74  & 0.2332 & 0.001 \\
\hline
\end{tabular}
\end{table}